\documentclass[letterpaper,dvipsnames]{article}
\usepackage{jheppub}
\usepackage{amsmath}
\usepackage{graphicx}% Include figure files
\usepackage{array}
\usepackage{ulem}
\usepackage{cancel}
\usepackage{xcolor}
\usepackage{slashed}
\usepackage{afterpage}
\usepackage{appendix}
\usepackage{multirow}
\usepackage{float}
\usepackage{mathtools}
\mathtoolsset{centercolon}

\usepackage{multirow}

\let\vec\mathbf
\newcommand{\ads}{{\rm AdS}}
\newcommand{\ds}{{\rm dS}}
\newcommand{\A}{\texttt{a}}
\newcommand{\B}{\texttt{b}}

\DeclareMathOperator{\csch}{csch}

\newcommand{\TeV}{\text{TeV}}

\newcommand{\uv}{{\rm uv}}
\newcommand{\ir}{{\rm ir}}

\newcommand{\es}[2] {\begin{equation} \label{#1} \begin{split} #2 \end{split} \end{equation}}
\newcommand{\D}{\mathrm{d}}
\newcommand{\mpl}{M_{\rm pl}}
\newcommand{\fdec}{f_{\rm dec}}

\title{Warped Dimensions at the Cosmological Collider}

\author[a]{Soubhik Kumar}
\author[b]{and Michael Nee}

\affiliation[a]{Institute of Cosmology, Department of Physics and Astronomy, Tufts University, Medford, MA 02155, USA}
\affiliation[b]{Department of Physics, Harvard University, Cambridge, MA 02138, USA}
\emailAdd{soubhik.kumar@tufts.edu}
\emailAdd{mnee@fas.harvard.edu}

\abstract{Extra dimensions are present in many beyond the Standard Model scenarios, most notably in string theory. However, direct signatures of extra dimensions are difficult to observe in many cases. This is the situation, for example, if the energy scales associated with extra dimensions are close to the string or Grand Unification scale. The energetic early universe provides an exciting opportunity to overcome this challenge, since the heavy states associated with high-scale extra dimensions, such as scalar moduli and Kaluza-Klein (KK) gravitons, could have been produced {\it on-shell} at early epochs. In this work, we illustrate this by focusing on how such states can be produced during inflation and leave signatures in primordial non-Gaussianity~(NG). Specifically, we consider a 5D spacetime with a warped extra dimension that remains stabilized as inflation proceeds in the four non-compact dimensions. By discussing an explicit stabilization mechanism, we compute the masses and couplings of the radion modulus and the KK graviton modes. Being gravitational degrees of freedom, these unavoidably couple to the field(s) generating curvature perturbation, and can lead to observable NG with a distinctive oscillatory shape and characteristic angular dependence. We give example benchmarks which can already be probed by the {\it Planck} data and identify targets for the future. Our study shows that cosmological surveys have the potential to observe on-shell imprints of extra dimensions in the coming years.}

\begin{document}

\maketitle

\section{Introduction}

\label{sec:intro}

Models with extra spatial dimensions are some of the most widely-studied scenarios of beyond the Standard Model (SM) physics. This stems from a theoretical interest in understanding quantum gravity and holography, as well as phenomenological motivations to explain observed features of the SM -- such as the hierarchy between the electroweak scale and the Planck scale~\cite{Arkani-Hamed:1998jmv, Antoniadis:1998ig, Arkani-Hamed:1998sfv, Randall:1999ee, Randall:1999vf}, and the hierarchical nature of the SM fermion masses~\cite{Grossman:1999ra, Gherghetta:2000qt, Huber:2000ie}. 

Extra dimensions also lead to robust predictions for new physics. The most observationally relevant predictions include new scalar moduli fields, deviations from Newton's law at small distances, and Kaluza-Klein (KK) towers of particles. Regardless of how the SM is embedded in the extra dimension, the existence of moduli fields, controlling the sizes of the extra dimensions, and the KK graviton modes is generic. Their properties are determined by two characteristic mass scales: the KK scale $\Lambda$ and the higher dimensional quantum gravity scale $\Lambda_{\rm QG}$ (assuming all the extra dimensions have a similar size). These scales are in turn set by the geometry of the extra dimension and the requirement to reproduce the 4D Newton's constant. Collider searches for KK resonances~\cite{CidVidal:2018eel, ATLAS:2024fdw, CMS:2024nht} and moduli~\cite{Giudice:2017fmj}, as well as laboratory tests of the gravitational force law~\cite{Murata:2014nra, Tan:2016vwu, Lee:2020zjt}, bound $\Lambda_{\rm QG}$ or $\Lambda$ to be greater than a few TeV, depending on the model.\footnote{In models with flat extra dimensions~\cite{Arkani-Hamed:1998jmv, Antoniadis:1998ig, Arkani-Hamed:1998sfv}, $\Lambda_{\rm QG} \gtrsim$~TeV, while the KK gravitons can appear at $\Lambda \ll$~TeV. On the other hand, in warped extra dimensional scenarios~\cite{Randall:1999ee, Randall:1999vf}, $\Lambda \gtrsim$~TeV, while $\Lambda_{\rm QG}\gg$~TeV.} This constrains extra dimensional models that aim to resolve the hierarchy problem~\cite{Arkani-Hamed:1998jmv, Antoniadis:1998ig, Arkani-Hamed:1998sfv, Randall:1999ee, Randall:1999vf}, but often the scale $\Lambda < \Lambda_{\rm QG}$ is predicted to be much higher (e.g. from string theory). In that case, most traditional experiments have little chance of directly observing the smoking-gun signatures associated with extra dimensions. Secondly, depending on the early universe dynamics, $\Lambda$ can be time-dependent. Thus, a scenario with a value of $\Lambda \sim \TeV$, necessary to solve the hierarchy problem, can have $\Lambda\gg$~TeV in the early universe.

Given that the energy scales in the early universe could have been much higher than the TeV scale, cosmological observables can provide a powerful complementary probe of physics at very high energy scales. Fields with masses of order the Hubble rate during cosmic inflation, $H$, can be produced via vacuum fluctuations and leave their imprint on cosmological correlators in the form of primordial non-Gaussianity~(NG)~\cite{Chen:2010xka, Wang:2013zva}. In this sense, the early universe acts as a `cosmological collider' with an extremely high energy reach~\cite{Chen:2009zp, Baumann:2011nk, Noumi:2012vr, Chen:2012ge, Arkani-Hamed:2015bza}, as $H$ may have been as high as $H\sim 10^{13}$ GeV~\cite{Planck:2018jri, ACT:2025tim}. For various ideas that explore the possible signatures of particle physics scenarios in the cosmological collider and advances in understanding and evaluating cosmological correlators themselves, see Refs.~\cite{Baumann:2011su, Assassi:2013gxa, Craig:2014rta,  Dimastrogiovanni:2015pla, Lee:2016vti, Meerburg:2016zdz, Chen:2016uwp, Chen:2016nrs, Chen:2016hrz, An:2017hlx, Chen:2017ryl, Kumar:2017ecc, Baumann:2017jvh, MoradinezhadDizgah:2018ssw, Goon:2018fyu, Chen:2018xck, Arkani-Hamed:2018kmz, Kumar:2018jxz, Wu:2018lmx, Dimastrogiovanni:2018uqy, Sleight:2019hfp, Lu:2019tjj, Hook:2019zxa, Hook:2019vcn, Kumar:2019ebj, Wang:2019gbi, Li:2019ves, Kim:2019wjo, Baumann:2019oyu, Alexander:2019vtb, Bodas:2020yho, Kogai:2020vzz,  Aoki:2020zbj, Lu:2021gso, Wang:2021qez, Lu:2021wxu, Cabass:2021fnw, Cabass:2021iii, Dimastrogiovanni:2021cif, Tong:2021wai, Cui:2021iie, Tong:2022cdz, Pimentel:2022fsc, Chen:2022vzh, Jazayeri:2022kjy, AnilKumar:2022flx, Maru:2022bhr, Qin:2022fbv, Cabass:2022jda, Niu:2022quw, Chen:2023txq, Qin:2023bjk, Chakraborty:2023qbp, Chakraborty:2023eoq, Aoki:2023dsl, Craig:2024qgy, Quintin:2024boj, Goldstein:2024bky, Qin:2024gtr, Hubisz:2024xnj, Liu:2024str, Stefanyszyn:2024msm, Bodas:2024hih, Cespedes:2025dnq, Chakraborty:2025myb, deRham:2025mjh, Zhang:2025nzd, Bodas:2025wuk, Qin:2025xct, Chakraborty:2025mhh, Aoki:2025uff}.

If the KK scale is of order the Hubble expansion rate during cosmic inflation, $\Lambda \sim H$, the moduli and KK gravitons can be gravitationally produced during inflation. Importantly, the high scale of inflation means that even very heavy moduli and KK gravitons could have been produced in such a fashion. Being gravitational degrees of freedom, these states generically couple to the field(s) responsible for sourcing density perturbations, so will leave their imprint on cosmological correlators in the form of NG.

It is therefore worth understanding the NG signatures of extra dimensions with $\Lambda \sim H$, since that might provide powerful clues regarding the structure of spacetime both in the early universe and at very small distances. In Ref.~\cite{Kumar:2018jxz}, one of us explored an inflationary scenario with a compact, flat extra dimension. It was shown that the inflationary vacuum energy gives rise to a horizon, and it is possible to have a near-horizon stabilization of the extra dimension such that its scale $\Lambda$ can indeed be $\sim H$. With such a setup, the NG signatures mediated by KK gravitons were evaluated and connections with Grand Unification were explored. However, due to the flat geometry of the extra dimension, the KK graviton couplings to the inflaton are suppressed by the 4D Planck scale ($M_{\rm pl}$), leading to a suppressed NG signal. Furthermore, stabilization of the extra-dimensions was not studied in full detail. As a result, the properties of the radion modulus, which controls the size of the extra dimension, could not be derived. Thus, the strength of the radion-mediated NG signature remained unknown.

In this work, we improve upon all these aspects. We focus on a warped extra dimension~\cite{Randall:1999ee, Randall:1999vf}, which can lead to a larger coupling between the extra-dimensional states and the fields responsible for generating the curvature perturbation. We stabilize the model by considering a Goldberger-Wise (GW) field~\cite{Goldberger:1999uk, Goldberger:1999un}, and explicitly show that it is possible to stabilize the extra dimension so that $\Lambda \sim H$. After incorporating the effects of the stabilizing sector, we find a tower of spin-2 KK gravitons with mass splitting $m_{\rm KK} \sim \Lambda \sim H$, and a radion, $\varphi$, whose mass is also set by $\Lambda$, $m_\varphi \sim\Lambda \sim H$. This setup leads to a spin-0 cosmological collider signal mediated by the radion, and a spin-2 signal generated by KK graviton modes. A simultaneous observation of a combined spin-0 and spin-2 NG signal would therefore provide a strong hint towards an extra-dimensional nature of the inflationary universe. 

The KK gravitons and radion couple most strongly to fields localized towards the infrared (IR) boundary. Therefore, the NG signals are largest when the field responsible for generating density perturbations is also localized in the IR. We will argue that the inflaton, however, cannot be an IR-localized field, as the energy densities produced by fields in the IR are too small to drive inflation if~$\Lambda \sim H$.\footnote{It is possible to have the inflaton as an IR-localized field if $\Lambda \gtrsim (H \mpl)^{1/2}$, in which case, however, the `non-local' NG signatures (characteristic of on-shell heavy particles) are negligible.} We are thus led to a scenario where the field driving the homogeneous inflationary expansion is localized on the ultraviolet (UV) boundary and is different from the field that generates density perturbations. This separation of roles is indeed the main feature of the curvaton mechanism~\cite{Lyth:2001nq, Moroi:2001ct, Enqvist:2001zp}. We find the NG signal is strongest in models with a UV-localized inflaton and an IR-localized curvaton.

Our setup has a dual interpretation via the AdS/CFT correspondence~\cite{Maldacena:1997re, Witten:1998qj, Gubser:1998bc}. In the field theory language, we are considering the confined phase of a strongly coupled CFT during inflation~\cite{Gubser:1999vj, Arkani-Hamed:2000ijo, Rattazzi:2000hs}. The KK gravitons are interpreted as spin-2 glueballs in the CFT, and their couplings are suppressed by powers of $\Lambda$ with coefficients determined by large-$N$ counting~\cite{tHooft:1973alw, Witten:1979kh}. This is in contrast to the $\mpl$ suppression in the case of a flat extra dimension~\cite{Han:1998sg} (as seen in~\cite{Kumar:2018jxz} in the context of NG) and allows for larger NG signals if the field responsible for the curvature perturbation is also a composite state. This is indeed the case if the curvaton is an an IR-localized field. The expansion history required in the curvaton scenario can then arise naturally if the SM is mostly elementary (i.e., localized on the UV boundary in the AdS picture). The lightest composite state will then be long-lived as it has only gravitational strength coupling to the fields on the UV boundary. In such a scenario, it is reasonable to assume that the curvaton decays much more slowly than the inflaton and comes to dominate the energy density before decaying. In this case, the curvaton fluctuations determine the density perturbations observed in the cosmic microwave background (CMB) and the large-scale structure (LSS).

Most observational searches for the cosmological collider signal have focused on the three-point function~\cite{Sohn:2024xzd, Cabass:2024wob, Goldstein:2024bky}, but in our model the dominant NG comes from the four-point function (trispectrum), as the three-point function is suppressed by the small curvaton mass. We find that both the spin-2 signal and the radion-mediated spin-0 signal are suppressed by the inverse powers of the large-$N$ parameter of the dual 4D theory and a `Boltzmann suppression' factor $\exp(-\pi m/H)$. We also find benchmark parameters where the spin-0 signal is at the boundary of the trispectrum constraints found in Ref.~\cite{Philcox:2025bvj, Philcox:2025lrr, Philcox:2025wts}, while the associated signal due to the lightest KK graviton is an order of magnitude smaller. The NG mediated by the radion and the KK gravitons could therefore be seen in upcoming surveys, highlighting the possibility of using NG to probe high-scale extra dimensions.

The rest of this paper is organized as follows. In section~\ref{sec:4dCFT} we discuss the 4D CFT picture of our setup, highlighting the cosmological history and providing general estimates for the relevant couplings. In section~\ref{sec:5dEFT} we discuss the 5D gravity picture, deriving the couplings of the radion and KK gravitons from a 5D construction. In section~\ref{sec:stabilisation} we incorporate a stabilization mechanism and numerically determine the properties of the radion and the spectrum of low-lying KK graviton states for two benchmark parameter points. In section~\ref{sec:NG} we calculate the trispectrum for these benchmark points and compare it to observational results. We conclude and summarize our results in section~\ref{sec:conclusions}.

\section{Composite Resonances During Inflation}

\label{sec:4dCFT}

The model we consider has an interpretation as the confined phase of a strongly coupled CFT via the AdS/CFT correspondence. In this section, we discuss this interpretation. This allows for a qualitative understanding of the theory before we use the 5D setup in section~\ref{sec:5dEFT} to derive precise expressions. The observability of the NG signal requires that there are fields with masses of order $H$ with sufficiently large couplings to the field that generates the curvature perturbations. We provide an outline of the cosmological scenario that leads to this being the case.

In the CFT picture, the inflaton, $\phi$, can either be a composite or an elementary field. If the inflaton is composite, then we expect the potential, $V(\phi)$, driving inflation to be of order the confinement scale (or smaller, if the inflaton is a pseudo Nambu-Goldstone boson (PNGB)), $V(\phi) \sim H^2 \mpl^2 \lesssim \Lambda^4$. This implies that the masses of composite states $\sim \Lambda$ are parametrically above the Hubble scale, as $\Lambda \gg H$. Thus, cosmological production of such states from vacuum fluctuations is exponentially suppressed, and their on-shell signatures are difficult to observe. The inflaton should therefore be elementary for the composite sector to be relevant during inflation, for which we require $\Lambda \simeq H$. As the couplings between elementary and composite fields are weak, it will be difficult to see the impacts of composite fields on inflationary observables in single field inflation.

However, the inflaton may not be the same field as the one giving rise to density fluctuations seen in the CMB and LSS~\cite{Lyth:2001nq, Moroi:2001ct, Enqvist:2001zp}. If there is a curvaton field, $\sigma$, that decays more slowly than $\phi$, then it can come to dominate the energy density after inflation has ended and be responsible for producing the observed density fluctuations. The curvaton can be composite and thus have $\mathcal{O}(1)$ couplings to other bound states. These bound states can then lead to a non-Gaussian spectrum of observed cosmological fluctuations. Such bound states include spin-0 states, such as the dilaton (dual to the radion), and a tower of spin-2 states, which lead to a distinctive angular dependence in the correlation functions. The spin-2 states are the glueballs of the CFT, which are described by the KK tower of gravitons in the 5D picture. 

The cosmological history of the scenario we have in mind is as follows. The inflationary expansion is driven by an elementary sector scalar field $\phi$ that provides the necessary potential energy density. During this inflationary epoch the CFT is in the confined phase, and $\sigma$ is misaligned on its potential.  In this case, the energy density of the curvaton can easily be smaller than $H^4 \lesssim \Lambda^4$, ensuring EFT control. We assume $\sigma$ is the lightest state of the composite sector, so it is natural to expect that the energy density in $\sigma$ is much less than the inflationary energy density and $\sigma$ acts as a spectator during inflation. Inflation ends when $\phi$ rolls sufficiently down its potential and starts oscillating, eventually decaying into SM radiation at a rate $\Gamma_\phi$. In the meantime, $\sigma$ starts oscillating in its potential when $H$ drops below its mass, and later decays into SM radiation with a rate $\Gamma_\sigma$. The end of this decay then marks the beginning of the standard radiation-dominated era in the early universe.

In the regime $\Gamma_\sigma \ll \Gamma_\phi$, $\sigma$ is much longer lived than $\phi$ and there can come to dominate the energy density before decaying into radiation. In this case the observed density fluctuations in the CMB and LSS arise from fluctuations in $\sigma$, rather than fluctuations in $\phi$. The reheat temperature from $\sigma$ decay, $T_{\rm RH} \sim \sqrt{\, \Gamma_\sigma \mpl}$, is below the confinement temperature if the curvaton is sufficiently long-lived, which means that throughout the cosmological history the composite degrees of freedom do not deconfine. Furthermore, the scattering rate of the elementary degrees of freedom with the composites is sub-Hubble, indicating negligible energy transfer between the sectors. Therefore, the composite sector does not deconfine via scattering either.

\subsection{Explaining the Current Observations} 

Here we show how this scenario can be consistent with CMB observations. In general, the total dimensionless curvature power spectrum ${\cal P}_\zeta$ is determined by both the inflaton and curvaton fluctuations. In terms of the inflaton power spectrum ${\cal P}_\phi$, and the isocurvature power spectrum ${\cal P}_{S}$ sourced by the curvaton fluctuations, ${\cal P}_\zeta$ is given by (see, e.g,~\cite{Lyth:2001nq, Fonseca:2012cj})
\es{eq:zetatot}{
    {\cal P}_\zeta = {\cal P}_{\phi} + {1 \over 9} \fdec^2 {\cal P}_{S} \, ,
}
where $\fdec$ is the fractional energy density in $\sigma$ at the time of its decay.\footnote{We note that ${\cal P}_{S}$ is the isocurvature perturbation during inflation, being sourced by the difference in fluctuations of the inflaton and the curvaton. Since the curvaton decays into SM fields in this set up, there is no surviving isocurvature perturbation during CMB decoupling.} We assume that the curvaton decays after the inflaton, the period between the two decays is radiation dominated, and that $\sigma$ oscillations start before the inflaton decays. With these assumptions $\fdec$ is given by\footnote{If $\fdec \sim 1$ then the universe will become matter dominated before $\sigma$ decays. The second equality in~\eqref{eq:f-decay} will therefore not be accurate but in this case $\fdec \simeq 1$ and the following analysis will still be valid.}
\es{eq:f-decay}{
    \fdec = {3\rho_\sigma \over 3\rho_\sigma + 4 \rho_{\rm r}}\bigg\rvert_{\sigma-{\rm decay}}  = {3 \sigma_i^2 \, \Gamma_{\phi}^{1/2} \over 4 \phi_{\rm osc}^2  \, \Gamma_{\sigma}^{1/2} + 3 \sigma_i^2 \, \Gamma_{\phi}^{1/2} }\, ,
}
where $\rho_\sigma$ is the curvaton energy density and $\rho_r$ is the energy density in radiation, and $\sigma_i, \ \phi_{\rm osc}$ are the field values of $\sigma, \ \phi$ when they start to oscillate. We have approximated the potentials for $\sigma$ and $\phi$ as quadratic potentials with no mixing of the fields. Requiring that the curvaton EFT is under control, implies that the curvaton energy density $m_\sigma^2 \sigma_i^2$ is subdominant compared to the compositeness scale $\Lambda^4$, which is a natural expectation if $\sigma$ is a PNGB, in which case $m_\sigma \lesssim \Lambda $. For similar models with an axion-like PNGB acting as a curvaton, see, e.g.~\cite{Dimopoulos:2003az,Kasuya:2009up, Kawasaki:2012wr, Mazumdar:2010sa}.

If both $\sigma$ and $\phi$ are only gravitationally coupled to the SM then $\Gamma_{\phi} \sim m_\phi^3 /\mpl^2$ and  $\Gamma_{\sigma} \sim m_\sigma^3 /\mpl^2$. This will be the case, for example, if the SM fields are elementary and the inflaton couples only gravitationally (by assumption). In this case $\fdec$ becomes
\es{}{
    \fdec \approx {3 \sigma_i^2 m_{\phi}^{3/2} \over 4 \phi_{\rm osc}^2 m_{\sigma}^{3/2} + 3 \sigma_i^2 m_{\phi}^{3/2} } \, .
}
The curvaton is long-lived relative to the inflaton if $m_\sigma \ll m_\phi$ and $\sigma_i \lesssim \phi_{\rm osc}$, in which case $\fdec \sim 1$.\footnote{Note, our above assumption that the oscillations of $\sigma$ start before the decay of $\phi$, is easy to satisfy since this imposes a weak restriction $m_\sigma \gtrsim m_\phi^3/\mpl^2$.} The inflaton may also have stronger couplings to the SM, in which case $\Gamma_\phi$ is larger, and the curvaton is more likely to dominate the energy density. An alternative possibility is that the SM fields are also bound states, in which case $\Gamma_\sigma \sim m_\sigma^3/\Lambda^2$. This then requires $m_\sigma$ to be much smaller, compared to the above, for the curvaton to be long-lived.

While it is not necessary for the validity of our results, to compare to cosmological observables we now consider scenarios with $\fdec \approx 1$ (for a recent discussion that tracks how $\fdec$ affects cosmological observables, see Ref.~\cite{Lodman:2023yrc}). For ${\cal P}_S \gg {\cal P}_\phi$, the primordial spectrum is then primarily determined by ${\cal P}_S \simeq H^2/(\pi\sigma_i)^2$, leading to
\es{eq:ps}{
    {\cal P}_\zeta \approx {H^2 \over 9 \pi^2\sigma_i^2}.
}
Thus $\sigma_i \approx 2.3 \times 10^3 H$ based on the inferred amplitude of primordial scalar perturbations~\cite{Planck:2018jri}.\footnote{It is possible to have a reduced hierarchy between $\sigma_i$ and $H$. From~\eqref{eq:zetatot}, if $\fdec\ll 1$, ${\cal P}_\sigma$, and hence, $H/\sigma_i$ can be larger, while keeping the combination $\fdec^2 {\cal P}_\sigma$ fixed. In that case $f_{\rm NL}^{\rm loc}$ increases $\propto 1/\fdec$~\cite{Lyth:2002my} and current constraints on $f_{\rm NL}^{\rm loc} = -0.9 \pm 5.1$~\cite{Planck:2019kim} become relevant.} The slow roll parameter $\epsilon$ is determined by the inflaton potential $\epsilon\equiv -\dot{H}/H^2 \approx m_\phi^2/(3H^2)$. 
This controls the spectral tilt of ${\cal P}_\zeta$,
\es{}{
    n_s - 1 \approx -{2\over 3}  {m_\phi^2 \over H^2}.
}
Recent measurements of $n_s$, using {\it Planck}, ACT, and DESI data~\cite{ACT:2025fju}, require $m_\phi^2 \approx 0.04H^2$. The ratio of the tensor power spectrum ${\cal P}_T$ and the scalar power spectrum is given by
\es{}{
    r = {{\cal P}_T \over {\cal P_\zeta}} \approx 18 {\sigma_i^2 \over\mpl^2},
}
which shows that for sufficiently small $\sigma_i$, the current upper limit $r<0.038$~\cite{ACT:2025tim} can be easily satisfied. The curvaton scenario also predicts an observable amount of local NG (bispectrum) with a strength~\cite{Lyth:2002my},
\es{}{
    f_{\rm NL}^{\rm loc} \approx -{5\over 4},
}
within the target sensitivity of the recently launched SPHEREx mission~\cite{SPHEREx:2014bgr}.

\subsection{Curvaton Couplings and Large-$N$ Counting}

\label{subsec:couplings}

The curvaton will have couplings to other bound states of the CFT, which can be parametrically estimated using large-$N$ counting. Of particular interest for us will be the coupling of $\sigma$ to the radion, $\varphi$, and spin-2 bound states which correspond to KK gravitons $\tilde h_{\mu \nu}$ in the 5D setup. The leading terms can be written as,
\begin{align}
    \mathcal{L} \sim \frac{c}{M} \tilde h_{\mu \nu }\nabla^\mu\sigma\nabla^\nu\sigma + \frac{c'}{M'}  \varphi \nabla^\mu\sigma\nabla_\mu\sigma 
    \, .
    \label{eq:CFT-coupling}
\end{align}
Other than the curvaton mass $m_\sigma$, the two independent scales in the theory are the Planck scale $\mpl$ and the confinement scale $\Lambda$, while the couplings $c, c'$ can contain additional factors of $N$. Since the interactions above should survive in the limit $\mpl \to \infty$, the scales $M, M'$ which appear in~\eqref{eq:CFT-coupling} must be identified with $\Lambda$. There are also couplings of the radion and KK modes to $\sigma$ that are proportional to $m_\sigma$. Due to the smallness of $m_\sigma/H$, these terms are suppressed and they mediate a smaller NG signal compared to the couplings via the kinetic terms, so are not shown explicitly.

Both the radion and KK gravitons can be interpreted as glueballs in the dual description of the theory. One way to see this is that their kinetic terms scale with $N^2$, characteristic of glueballs. The glueball 3-point function scales like $1/N$ in the 't Hooft limit~\cite{Witten:1979kh}, implying that $c, c' \sim 1/N$. This again aligns with the gravitational picture (section~\ref{sec:5dEFT}), where the kinetic terms for $\tilde h_{\mu \nu}$ and $\varphi$ are schematically given by,
\begin{align}
    \mathcal{L} \sim N^2 \left( \nabla^\mu \tilde{h}^{\nu \rho}\nabla_\mu \tilde{h}_{\nu \rho} + \nabla^\mu \varphi \nabla_\mu \varphi \right) \, .
\end{align}
After canonically normalizing, the operators in \eqref{eq:CFT-coupling} will be multiplied by factors of $1/N$. 

\section{5D Gravitational Theory}

\label{sec:5dEFT}

In this section, we describe the 5D gravitational theory. The geometry is a slice of $\ads_5$ spacetime bounded by two branes, the UV and the IR branes (boundaries). These boundaries contain localized energy densities that give rise to inflationary expansion of 4D slices~\cite{Csaki:1999mp, Csaki:1999jh, Binetruy:1999hy}. Similar models with inflation generated by brane-localized fields have been considered extensively in the past~\cite{Lukas:1998qs, Lukas:1999yn, Nihei:1999mt, Maartens:1999hf, Karch:2020iit, Hubisz:2024xnj}. In the context of an $\ads_5$ geometry, most studies have focused on the case where the IR brane is behind the cosmological horizon, $\Lambda \to 0$. The theory is then a gapped CFT with a continuum of states above the scale $H$, a situation which has also been studied from the purely 4D perspective and been shown to lead to non-Gaussian density perturbations~\cite{Green:2013rd, Aoki:2023tjm, Pimentel:2025rds}. Another well-studied scenario where extra dimensions are invoked is brane inflation, where a brane moves through a warped background, which has an inflationary metric induced on the brane~\cite{Dvali:1998pa, Kaloper:1999sm,Arkani-Hamed:1999fet, Burgess:2001fx, Garcia-Bellido:2001lbk, Jones:2002cv, Kachru:2003sx, Silverstein:2003hf, Alishahiha:2004eh}. In this work, we instead consider situations where the IR brane is present, and the UV-IR brane separation, parametrized by the vacuum expectation value (VEV) of the radion field, remains constant during inflation.

\subsection{5D Geometry}

The 5D action is given by 
\es{eq:5daction}{
	S = \int \D^4 x \int_{-L}^L \D y \sqrt{-G}(M_5^3 {\cal R}_5-\Lambda_5) + \int \D^4 x \int_{-L}^L \D y \sqrt{-G}\left(-{1\over 2}G^{MN}\partial_M\Phi \partial_N\Phi - V(\Phi) \right)\\
	+\int \D^4 x \int_{-L}^L \D y \sqrt{-G} \left[ \left( {\cal L}_\phi - V_0 (\Phi) \right) \delta(y)+ \left( {\cal L}_\sigma - V_L (\Phi) \right) \delta(y-L) \right] ,
}
with the AdS bulk cosmological constant $\Lambda_5 = -12 M_5^3 k^2$ and 5D Ricci scalar ${\cal R}_5$. Here $M_5$ is the 5D Planck scale and $k$ is the AdS curvature scale. The second term in~\eqref{eq:5daction} denotes the action of a Goldberger-Wise scalar field $\Phi$ that will be relevant for the stabilization of the extra dimension~\cite{Goldberger:1999uk, Goldberger:1999un}. The third term in~\eqref{eq:5daction} captures boundary-localized terms, which include the inflaton and curvaton Lagrangians, denoted ${\cal L}_\phi$ and ${\cal L}_\sigma$ respectively.

The homogeneous 5D metric is given by,
\es{eq:metric_y}{
	\D s^2 = G_{MN}\D x^M \D x^N = n(y)^2 g_{\mu \nu}\D x^\mu \D x^\nu + \D y^2.
}
Here and below, we will use $G_{MN}$ and $g_{\mu\nu}$ to denote the bulk 5D metric and the 4D metric induced on a constant $y$ hypersurface, respectively. The static  Randall-Sundrum (RS) geometry~\cite{Randall:1999ee, Randall:1999vf}, in the absence of backreaction from $\Phi$, is obtained for a warp factor $n(y)=\exp(-ky)$ and $g_{\mu \nu} = \eta_{\mu \nu}$ (Minkowski metric), with $V_0 = 12M_5^3 k$ and $V_L = -12M_5^3k$.

Our goal is to study how this geometry is changed when $V_0$ and $V_L$ take on different values such that the 4D slices undergo inflationary expansion with Hubble rate $H$, in which case the 4D metric is given by:
\begin{align}
    &g_{\mu \nu}\D x^\mu \D x^\nu = -\D t^2 + a(t)^2 \D \vec{x}^2  = {1\over \eta^2 H^2} (-\D \eta^2 + \D \vec{x}^2)\, ,
    &&a(t) = \exp(H t) \, .
    \label{eq:4dmetric}
\end{align}
Here we have also expressed the metric in terms of conformal time $\eta$, which we will use to compute NG in the later sections.

Using the stress-energy tensor of the GW field, the $00$ and $55$ components of the Einstein equations can be written as~\cite{DeWolfe:1999cp},
\es{eq:G00}{
	G_{00} = 3 (H^2-n'^2-nn'') = {1\over 2 M_5^3} T_{00} = {n^2\over 2 M_5^3} \left(\Lambda_5 +{1\over 2} \Phi'^2 + V\right) + {n^2\over 2 M_5^3}\left(V_0 \delta(y)+V_L\delta(y-L)\right),
}
\es{eq:G55}{
	G_{55} = {6\over n^2}(n'^2 - H^2) = {1\over 2 M_5^3} T_{55} = {1\over 2 M_5^3} \left(-\Lambda_5 + {1\over 2} \Phi'^2 - V\right)\, .
}
The boundary potential energies determine the junction conditions on the warp factor, with $\epsilon$ infinitesimally positive: 
\begin{align}\label{eq:bc}
	&\left[\frac{n'}{n}\right]_{-\epsilon}^{+\epsilon} = - \frac{V_0}{6 M_5^3},
    &&\left[\frac{n'}{n}\right]_{L+\epsilon}^{L-\epsilon} = \frac{V_L}{6 M_5^3}.
\end{align}
The equations~\eqref{eq:G00},~\eqref{eq:G55},~\eqref{eq:bc} hold for arbitrary warp factors $n(y)$.
However, we can obtain analytical solutions and gain insight by first dropping the GW contribution.

\subsection{Geometry in the Absence of Stabilization}

\label{subsec:geometry}

Here we discuss the solutions for the warp factor and KK spectrum when $\Phi = 0$. In section~\ref{sec:stabilisation}, we will include $\Phi$ and discuss how to stabilize the extra dimension, but many of the results can already be obtained by considering the model for $\Phi=0$. In fact, for the example benchmarks we end up considering, we will see that the leading order spacetime geometry approximately follows the $\Phi=0$ solution, except in a small region near the IR boundary.

The bulk equations of motion (EOM)~\eqref{eq:G00} and~\eqref{eq:G55} can be simplified as $n'' = n k^2$, which is solved by
\es{}{
	n(y) = c_1 e^{ky} + c_2 e^{-ky} \, ,
}
where the coefficients $c_1$ and $c_2$ are determined by the boundary conditions~\eqref{eq:bc}. As discussed in the previous section, we consider an inflaton localized to the UV brane. For now we take the inflaton potential, $V_{\rm inf}$, to be constant, leading to exact de Sitter space. This contribution modifies the boundary tension in the UV to be $V_0 = 12 M_5^3k + V_{\rm inf}$.

We also allow for a detuned brane tension in the IR, $V_L = -12 M_5^3k - \tilde{V}$.\footnote{The curvaton energy density will in general contribute to $V_L$, but is much smaller than the critical detuning so we can ignore it's contribution.} The conditions satisfied by $c_{1,2}$ are then
\begin{align}
	&c_1 + c_2 = 1,
    &&c_1 - c_2 = -(c_1+c_2)\left(1 + {V_{\rm inf} \over 12 M_5^3k}\right),
\end{align}
where we have imposed a normalization $n(0)=1$ and the second expression is the boundary condition at $y=0$. The solutions are given by $c_1 = -V_{\rm inf}/(24 M_5^3 k)$ and $c_2 = 1+ V_{\rm inf}/(24 M_5^3 k)$. Plugging the expression of $n(y)$ into eq.~\eqref{eq:bc}, with $c_{1,2}$ derived above, we obtain the relation between $V_{\rm inf}$ and $H$:
\es{eq:Vinf_sol}{
	V_{\rm inf} = 12 M_5^3\left(\sqrt{k^2+H^2}-k\right).
}
Using this relation, we can express $c_{1,2}$ in terms of $H$ and $k$, and obtain an alternate form of the warp factor,
\es{eq:n_sol}{
	n(y) = \cosh(ky) - {\sqrt{k^2+H^2}\over k}\sinh(ky).
}
To satisfy the jump condition on $y=L$ boundary, we need $\tilde{V}\neq 0$. Defining 
\begin{align}\label{eq:alpha_beta}
    &\alpha \equiv {V_{\rm inf} \over 12M_5^3 k},
    &&\beta \equiv {\tilde{V}\over 12M_5^3 k},
\end{align}
as the two detuning parameters that parametrize the deviation from the original RS conditions~\cite{Randall:1999ee}, we find
\es{eq:size}{
	\exp(2 k L) = {\beta(2+\alpha) \over \alpha (2+\beta)}.
}
For $0<\alpha \ll 1$, $0 < \beta \ll 1$, such that the amount of detuning is small, the above equation has a solution only when $\beta > \alpha$. In this case we can approximate
\es{eq:approx_size}{
	\exp(2 k L) \approx {\beta \over \alpha},
}
and a large warp factor is obtained for $\beta \gg \alpha$. As evident from the form of the warp factor in~\eqref{eq:n_sol}, the geometry features a horizon. To derive its location, we first express
\es{eq:alpha}{
	\alpha = \sqrt{1+{H^2 \over k^2}} - 1 \, .
}
The location of the horizon is given by $n(y_H)=0$ with 
\es{}{
	\exp(2 k y_H) = 1 + {2\over \alpha}.
}
The horizon is behind the IR boundary as long as $L < y_H$ which gives $\beta/(2+\beta) < 1$ and $\alpha+2>0$, both of which are obeyed given the conditions discussed above~\eqref{eq:approx_size}. Note the above conditions do not necessarily imply that the extra dimensional geometry is stable. In fact, we will see below that the radion fluctuation around this solution is tachyonic, indicating an instability~\cite{Chacko:2001em}. The solution~\eqref{eq:approx_size} is a maximum of the radion potential; to find a stable solution we need to introduce a stabilization mechanism, which we will consider in the following section.

\subsection{Conditions for 5D EFT Control} 

Within this setup, several conditions are required to have a viable 5D EFT description. For 5D EFT control, we first require that the AdS curvature is sub-Planckian, $k < M_5$. The warped down 5D Planck scale and the Kaluza-Klein scales are given by $\Lambda_c \equiv n(L) M_5$ and $\Lambda\equiv n(L) k$, respectively, where $n(L)$ is the warp factor on the IR brane. This relation makes precise what we meant by $\Lambda$ in the above discussions. On the IR boundary, we then have a controlled 5D description for energies $\Lambda < E <  \Lambda_c$, while for energies  $E < \Lambda$ the theory is effectively 4D. 
This requires a hierarchy between $k$ and $M_5$, which we parametrize by the large-$N$ parameter 
\begin{align}
     N \equiv \left( \frac{M_5}{k} \right)^{3/2}\, .
    \label{eq:largeN}
\end{align}

We now revisit the motivation for considering the curvaton as a composite field from the 5D perspective.
First, we see why it is problematic to treat the inflaton as composite -- which corresponds to localizing the inflaton on the IR boundary in the gravitational picture. For EFT control, the inflaton energy density should be smaller than the warped down 5D gravity scale $\Lambda_c$, which requires: 
\es{}{
    {V_{\rm inf} \over \Lambda_c^4} \sim {H^2 M_5^3 \over k \Lambda_c^4} \sim {H^2 \over \Lambda^2} {1\over n(L)^2}{1\over N^{2/3}} \ll 1.
}
We are interested in $\Lambda\simeq H$ such that the cosmological production of the KK graviton states is not exponentially suppressed. Additionally, we require $n(L)\ll 1$, such that the coupling of the KK gravitons to the curvature perturbations is only suppressed by $ \Lambda_c = n(L) \mpl \ll \mpl$. In particular, the benchmark examples we discuss in the next section have $n(L) \approx 10^{-4}$. Thus, we see that unless $N$ is extremely large, we are led to $V_{\rm inf} \gg \Lambda_c^4$, violating the above condition and implying that the 5D EFT is not under control in this regime. This points to the inflaton being on the UV boundary. However, then its coupling to the graviton KK modes is suppressed by $\mpl$ since those KK modes are localized in the IR, with negligible overlap on the UV boundary.

As discussed in the previous section, a curvaton localized on the IR brane resolves this issue. The graviton KK modes will then couple significantly to the curvaton and lead to non-Gaussianities, while remaining under EFT control. In that case, comparing the energy densities on the IR boundary gives us
\es{}{
    {V(\sigma)\over \Lambda_c^4} \sim {m_\sigma^2 \sigma_i^2 \over \Lambda_c^4} \sim {m_\sigma^2 \over H^2} {\sigma_i^2 \over H^2} {H^4 \over \Lambda_c^4}.
}
As seen from~\eqref{eq:ps}, explaining the density perturbations requires $\sigma_i \gg H$. However, a small curvaton mass $m_\sigma \ll H$ can naturally compensate this to ensure $V(\sigma) \ll \Lambda_c^4$, indicating 5D EFT control.

\subsection{KK Graviton Fluctuations} 

Having discussed the behavior of the homogeneous metric and the conditions for a valid 5D EFT, we now study the tensor fluctuations around the background solution. The tensor modes can be parametrized as $G_{\mu \nu} = n^2(y) \bar g_{\mu \nu}+ h_{\mu\nu}(x,y)$, where $\bar g_{\mu \nu}$ is the background de Sitter solution~\eqref{eq:4dmetric} and $h_{\mu\nu}$ satisfies the transverse $\nabla_\mu h^{\mu\nu}=0$ and traceless $h_\mu^\mu=0$ constraints. The full metric then becomes
\es{eq:metric}{
	\D s^2 = n(y)^2(-\D t^2 + a(t)^2 \D \vec{x}^2) +\D y^2 + h_{\mu\nu}(x,y)\D x^\mu \D x^\nu,
}
At linear order, the Einstein equations in terms of $h_{\mu\nu}$ reduce to (see~\cite{Kumar:2018jxz} for a derivation),
\es{}{
	\square_{\ds}h_{\mu\nu} +n^2 h_{\mu\nu}'' -2 n'^2 h_{\mu\nu}-2nn'' h_{\mu\nu}-2 H^2 h_{\mu\nu}=0 \, ,
}
where the Laplacian $\square_{\ds}$ is computed using the 4D $\ds$ metric $\bar g_{\mu \nu}$. We can then do a KK decomposition,
\es{eq:KK}{
	h_{\mu\nu}(x, y) = \sum_l n^2(y) \tilde{h}_{\mu\nu,l}(x)\chi_l(y) \, ,
}
where 
\es{eq:htildemunu}{
    \square_{\ds}\tilde{h}_{\mu\nu,l}=(m_l^2+2H^2)\tilde{h}_{\mu\nu,l} \, ,
}
and
\es{eq:profile}{
	n^2\chi_l'' +4 n n'\chi'_l+m_l^2 \chi_l = 0 \, .
}
For the zero mode $m_l=0$ and the above equation can be solved for any warp factor $n(y)$:
\es{}{
	\chi_0'(y) = {C \over n^4}.
}
The $\chi_l$ satisfy Neumann boundary conditions, $\chi_l'(y)=0$ for $y=0, L$, which forces $C=0$ and $\chi_0(y)$ to be a constant.

To study the mode functions for the massive KK gravitons, it is convenient to go to conformal coordinates $z$ by demanding $\D y = n(y) \D z$ so that the homogeneous metric becomes
\es{eq:metric_z}{
	\D s^2 = f(z)^2 \left(-\D t^2 + a(t)^2 \D\vec{x}^2 + \D z^2\right),
}
with $n(y) = f(z)$. Defining $\chi_l(y) = \psi_l(z)/n^{3/2}$ with ${\D/\D y} = (1/n)(\D / \D z)$, and denoting $\dot{\psi}_l \equiv \D \psi_l/\D z$, the profile equation~\eqref{eq:profile} in terms of $\psi_l$ becomes,
\es{eq:eigen}{
	{\D^2\psi_l \over \D z^2} - {3\over 2}{\ddot{f} \over f}\psi_l - {3\over 4}{\dot{f}^2 \over f^2}\psi_l + m_l^2 \psi_l = 0.
}
When the GW field is absent, eqs.~\eqref{eq:G00} and~\eqref{eq:G55} determine $f(z) = H/(k \sinh(H z))$ and give the eigenvalue equation
\es{eq:KK_wavefn_1}{
	{\D^2\psi_l \over \D z^2} - {3\over 8}H^2(7+3\cosh(2 H z))\csch(H z)^2 \psi_l(z) + m_l^2 \psi_l(z) = 0,\\
    \Rightarrow {\D^2\psi_l \over \D \bar{z}^2} - {15\over 4}\csch(\bar{z})^2 \psi_l(\bar{z}) + \left({m_l^2\over H^2}-{9\over 4}\right) \psi_l(\bar{z}) = 0,
}
with $\bar{z}=zH$. This is now in the form of a Sturm-Liouville problem and thus the KK graviton masses satisfy $m_l^2 \geq 9H^2/4$. The KK graviton spectrum is therefore gapped from the zero mode, with a lower bound on the gap set by $3H/2$. In particular, these massive spin-2 particles satisfy the Higuchi bound~\cite{Higuchi:1986py}.

To determine the normalization condition for the KK graviton mode functions, one can expand the bulk action~\eqref{eq:5daction} to quadratic order in the tensor fluctuations and canonically normalize the 4D graviton. Going back to a generic warp factor, this gives, 
\es{eq:normalisation}{
    2 M_5^3\int_{z_{\uv}}^{z_\ir} \D z \psi_{l_1}^*(z) \psi_{l_2}(z) = 
    2 M_5^3\int_0^{L} \D y n^2(y) \chi_{l_1}^*(y) \chi_{l_2}(y) =  {M_4^2 \over 2}\delta_{l_1 l_2},
}
where $z_{\uv}$ and $z_\ir$ are the locations of the UV and the IR boundaries, respectively, in the $z$-coordinates. As $H\rightarrow 0$, the quantity $M_4 \rightarrow \mpl$, the reduced 4D Planck scale. For the zero mode, since $\chi_0(y)$ is a constant, we can set $\chi_0(L)=1$ and get
\es{}{
    M_4^2 = 4 M_5^3 \int_0^L \D y n^2.
}

\subsection{Radion Fluctuation}

The radion field controls the size the extra dimension, and in this case, the separation between the IR and the UV boundary. We parametrize the radion field $r(x)$ such that its VEV determines $L$ via $L = \pi \langle r(x) \rangle$, 
\es{eq:rad_ansatz}{
    \D s^2 = n^2\left(\frac{\pi  r(x)  y}{L}\right) (-\D t^2 + a(t)^2 \D \vec{x}^2) + \left(\frac{\pi r(x)}{L}\right)^2\D y^2.
}
In the discussion below we will drop the argument in $r(x)$, using $r$ to refer to the full field and using $L$ to refer to the VEV, where relevant. Using this ansatz, we can split the 5D action~\eqref{eq:5daction} into potential and kinetic terms for the radion:
\es{eq:radEFT}{
    &{\cal L}_{\rm rad} = {\cal L}_{\rm kin } - V_{\rm rad} \, .
}
After performing some algebra we derive the kinetic term for the radion as,
\es{eq:Lrad_kin}{
    {\cal L}_{\rm kin} = 12 M_5^3 \sqrt{-g} {(\partial_\mu r)^2 \over r^2}\int_{0}^{\pi r}\D y y n'(y) (n + y n'(y)) \, ,
}
where we note that ${\cal L}_{\rm kin}$ contains derivative interactions as well as the usual kinetic term. 

We can canonically normalize the radion kinetic term by expanding $r$ around its VEV, $r = \langle r \rangle +\delta r$, in ${\cal L}_{\rm kin}$~\eqref{eq:Lrad_kin}:
\es{}{
    {\cal L}_{\rm kin} = 12\pi^2 M_5^3 \sqrt{-\bar{g}} {(\partial_\mu \delta r)^2 \over L^2}\int_{0}^{L}\D y y n'(y) (n + y n'(y)) + \ldots \, , 
}
where the `$\ldots$' refer to derivative interactions of $r$ which we neglect. We can then define the canonically normalized radion field 
\begin{align}
    \varphi = F_\varphi e^{-k\pi r} \, ,
\end{align}
where
\es{eq:F_vaphi}{
   F_\varphi^2 = -\frac{24 M_5^3 e^{2k L}}{(k L)^2 } \int_{0}^{L}\D y \, y n'(y) (n + y n'(y))  \, .
}
This leads to the leading order kinetic term for the fluctuations, $\delta\varphi = F_\varphi e^{-kL }(-k\pi \delta r)$,
\begin{align}
    {\cal L}_{\rm kin} = -{1\over 2} \sqrt{- \bar g}  (\partial_\mu \delta \varphi)^2 + \ldots\, .
\end{align}
In the following section we will use the full expression~\eqref{eq:F_vaphi} to calculate $F_{\varphi}$ with a numerical solution for $n(y)$, but we can perform a sanity check by using $n=\exp(-ky)$, valid in the zero backreaction and $H=0$ limit. The integral then gives $(-1/2)k (\pi r)^2\exp(-2k\pi r)$, leading to
\es{}{
    \frac{{\cal L}_{\rm kin}}{\sqrt{-g}} \rightarrow -6 M_5^3 k\pi^2 \exp(-2k\pi r) {(\partial_\mu r)^2} = -{6 M_5^3 \over k}\left(\partial_\mu \exp(-k\pi r)\right)^2  .
}
This matches with the result of Ref.~\cite{Chacko:2013dra} (up to the different normalization convention for the 5D action).

Next we focus on the potential terms, using the dimension reduction formula ${{\cal R}^{(5)} = (1/n^2){\cal R}^{(4)}+\cdots}$ and the Einstein equations~\eqref{eq:G00} and~\eqref{eq:G55} to derive the radion potential from the bulk action~\eqref{eq:5daction}. Doing so gives
\es{eq:Vrad}{
    \frac{V_{\rm rad}}{\sqrt{-g} } =& 
    12  M_5^3 \left( n^3 (\epsilon) n'(\epsilon)- n^3 (\pi r-\epsilon) n'(\pi r-\epsilon) + \left( H^2 - \frac{{\cal R}_4}{6} \right) \int_0^{\pi r} \D y n(y)^2    \right)  
    + V_0 n(0)^4 + V_L n(\pi r)^4 
    \, .
}
To derive the above `off-shell' potential, we have not used the junction conditions~\eqref{eq:bc}, which are only satisfied at the extrema. We can again perform a sanity check in the zero backreaction and $H=0$ limit. Using $n(y) = e^{-k y}$ eq.~\eqref{eq:Vrad} becomes
\es{}{
    V_{\rm rad} \rightarrow 12 M_5^3 k\left(e^{-4 k \pi r} - 1\right) + V_0  + V_L e^{-4k \pi r}.
}
Therefore, the tuned values $V_0 = 12 M_5^3 k$ and $V_L = -12M_5^3 k$ lead to a vanishing radion potential, as expected. In general, the boundary conditions~\eqref{eq:bc} imply the $V_0 \, (V_L)$ terms and $n'(\epsilon) \, (n'(\pi r-\epsilon))$ terms in~\eqref{eq:Vrad} cancel each other at the extrema. Then the 4D cosmological constant comes from the $H^2$ term and is given to leading order by $6M_5^3 H^2/ k = 3 H^2 \mpl^2$, giving the correct vacuum energy that drives the inflationary expansion in the 4D EFT.\footnote{Note, the Ricci term ${\cal R}_4$ is part of the standard 4D Einstein-Hilbert action and does not contribute to 4D vacuum energy.}

With the generic radion potential~\eqref{eq:Vrad} at hand, we can also compute the radion mass in the unstabilized ($\Phi = 0$) model. To do so we take the solution~\eqref{eq:n_sol} for $n(y)$ and set ${\cal R}_4 = 12 H^2$, as the 4D homogeneous metric is exact de-Sitter space. The mass of the radion, to leading order in $(H/k)^2$, is then given by
\begin{align}
    m_{\varphi}^2 = \frac{\D^2 \, V_{\rm rad}}{\D \varphi^2} = \frac{1}{F_\varphi^2}\frac{\D^2 \, V_{\rm rad}}{\D (e^{-k \pi r})^2} = {24 H^2 (1 + \beta) \over -6 (1 + \beta) + \beta (2 + \beta) \log[((2 + \alpha) \beta)/(\alpha (2 + \beta))]} \rightarrow -4H^2.
    \label{eq:tachyon_mass}
\end{align}
where the expression has been evaluated at the critical point $\exp(2k\pi r) = \beta(2+\alpha)/(\alpha (2+\beta))$ (eq.~\eqref{eq:size}), with $\alpha, \, \beta$ defined in eq.~\eqref{eq:alpha_beta}. The final step is obtained assuming $\alpha,\beta\ll 1$ and the logarithmic factor is not large. Here we see explicitly that the radion is tachyonic in de Sitter space without a stabilization mechanism.

\subsection{Couplings to the Curvaton}

We can now determine the coupling of the curvaton to the radion and KK graviton fluctuations from the kinetic term for $\sigma$,
\begin{align}
    \mathcal{L}_{\rm kin, \, \sigma} &=  -{1\over 2} \sqrt{-G}G_{\mu\nu} \nabla^\mu \sigma \nabla^\nu\sigma 
    = - {1\over 2}  \sqrt{-\bar{g}} \left( \bar g_{\mu\nu}+  \sum_l \chi_l(L) \, \tilde{h}_{\mu\nu,l}\right) n^2(\pi r) \nabla^\mu \sigma \nabla^\nu\sigma \, .
    \label{eq:curv_kin}
\end{align}
We can then canonically normalize $\sigma$ by rescaling $\sigma \rightarrow \sigma/n(L)$ and the KK graviton by $\tilde{h}_{\mu\nu,l} \rightarrow \tilde{h}_{\mu\nu,l}/M_4$. Expanding around the radion background, $\pi r = L + \pi \delta r$, we then find the kinetic term and leading order couplings to the radion and KK gravitons
\es{eq:LsigmaKK}{
    \mathcal{L}_{\rm kin, \, \sigma} &= 
    - {1\over 2}  \sqrt{-\bar{g}} \left[ \bar g_{\mu\nu} \left(1 +  {2n'(L) \over n(L)} \pi \delta r\right)+ \frac{1}{M_4}\sum_l \chi_l(L) \, \tilde{h}_{\mu\nu,l}\right]   \nabla^\mu \sigma \nabla^\nu\sigma 
}
Since $\chi_0(L)=1$, this shows the curvaton couples to the graviton zero mode with $M_4$ suppression, as expected. However, for the KK modes, the graviton wavefunction is peaked at the IR boundary, which enhances the coupling to the curvaton. For example, the ratio of the coupling strength of the $l$-th KK mode with respect to the zero mode is given by, $\chi_l(L)/\chi_0(L) = \psi_l(z_{\rm ir})/\psi_0(z_{\rm ir}) = \psi_l(z_{\rm ir})/f^{3/2}(z_{\rm ir})$. With $\psi_l$ determined by solving~\eqref{eq:eigen}, the above ratio can be calculated for any warp factor.

We can then write the radion interaction term in $\mathcal{L}_{\rm kin, \, \sigma} $ in terms of the canonically normalized radion, $\varphi$, as
\es{eq:rad_couple_1}{
    {1\over 2} \left( {2n'(L) \over k n(L)} {\delta\varphi \over \langle \varphi\rangle}\right) \sqrt{-\bar{g}} \bar{g}^{\mu\nu}\nabla_\mu\sigma\nabla_\nu\sigma 
    =  { \lambda_{\varphi \sigma} \over 2} \frac{\delta\varphi}{\langle \varphi\rangle} \sqrt{-\bar{g}} \bar{g}^{\mu\nu}\nabla_\mu\sigma\nabla_\nu\sigma \, ,
}
where we have defined the radion-curvaton coupling (using $\Lambda = kn(L)$)
\es{eq:rad_couple_2}{
    \lambda_{\varphi \sigma}= {2n'(L) \over \Lambda}\, .
}

The discussion so far has assumed that there is no mixing between the curvaton and the radion. This mixing could be induced if there is a coupling of the curvaton to the Ricci scalar, $\xi \sigma^2 {\cal R}_4$~\cite{Giudice:2000av, Csaki:2000zn}. The coupling $\xi$ also generates a Hubble-scale mass for the curvaton, so must be small for $\sigma$ perturbations to generate an approximately scale-invariant power spectrum. For the curvaton scenario to make sense, we therefore require $\xi \ll 1$. Furthermore, if $\xi$ were nonzero there would be mixing set by angle $\tan(\theta) = 6\xi \langle\sigma\rangle/\langle\varphi\rangle$. Requiring this mixing angle to be small imposes a more stringent requirement $\xi \ll \langle\varphi\rangle/\langle\sigma\rangle$, since $\langle \sigma\rangle \sim \sigma_i \gg H \sim \langle\varphi\rangle$ to get the correct amplitude for the power spectrum (see eq.~\eqref{eq:ps}). Additionally, for $\xi \neq 0$, the coupling in~\eqref{eq:rad_couple_2} would be multiplied by an overall factor of $(1-6\xi)$. In the following, we assume $\xi$ is sufficiently small such that $\tan(\theta)\ll 1$, satisfying the more stringent requirement. However, it is straightforward to include the mixing effects if $\langle\varphi\rangle/\langle\sigma\rangle < \xi\ll 1$.

\section{Stabilization and the KK Graviton Spectrum}

\label{sec:stabilisation}
In this section we include the effect of a GW field $\Phi$, in order to stabilize the extra dimension. We will consider only a mass term for $\Phi$ in the bulk so that the equation of motion and boundary conditions for $\Phi$ can be solved analytically. We then use this solution to numerically solve for the backreaction of $\Phi$ on the metric, and with these ingredients derive the effective potential for the radion. Finally, using the numerical profiles for two benchmark parameter points, we derive the quantities relevant for calculating the NG signal. These quantities are: the radion mass $m_\varphi$, the radion VEV $\langle \varphi \rangle$, $F_\varphi$, the radion-curvaton coupling $\lambda_{\varphi \sigma}$, the IR scale $\Lambda$, and the lowest order KK graviton masses and wavefunctions. These results are summarized in table~\ref{tab:results} and the field profiles shown in figures~\ref{fig:profiles}~\&~\ref{fig:KKspectrum}.

\subsection{Equations of Motion}

\label{subsec:EOMs}

In order to source the GW field we include $\Phi$-dependent potentials on the boundaries in addition to the constant tensions (see eq.~\eqref{eq:5daction}):
\begin{align}
    &V_0 = 12 M_5^3k + V_{\rm inf}  + v_0 (\Phi) \, ,
    &&V_L =  -12 M_5^3k - \tilde{V} + v_L (\Phi) \, .
    \label{eq:bdy_potentials}
\end{align}
The background solution for $\Phi$ then satisfies the boundary conditions
\begin{align}\label{eq:phi_bc}
	&\Phi' \rvert_{-\epsilon}^{+\epsilon} = v'_0(\Phi) \, ,
    &&\Phi' \rvert_{L-\epsilon}^{L+\epsilon} =v'_L(\Phi) \, .
\end{align}
We also consider a bulk mass term for $\Phi$, so that the potential is given by
\begin{align}
    V(\Phi) =  \frac{\varepsilon k^2}{2} \Phi^2 \, ,
\end{align}
where $\varepsilon$ is a dimensionless parameter that controls the bulk mass of $\Phi$ in units of $k$.

As discussed in the previous section, the 4D effective potential reduces to a sum of boundary terms plus the term generated by $H$:
\begin{align} \label{eq:Veff}
    {V_{\rm eff}\over \sqrt{-\bar{g}}} =& n^4(\pi r) \left[ -12 M_5^3 \frac{n'(\pi r)}{n(\pi r)} - 12 M_5^3k - \tilde{V} + v_L (\Phi(\pi r))\right]
    \nonumber \\
    &+  \left[ 12 M_5^3k + V_{\rm inf} + v_0 (\Phi(0)) +12 M_5^3 n'(0) \right]
    - 12 M_5^3 H^2 \int_0^{\pi r} \D y n(y)^2 \, ,
\end{align}
where we have used the normalization condition $n(0)=1$. We have also set $\mathcal{R}_4 = 12 H^2$ to include the $\mathcal{R}_4$ term in addition to the $H^2$ term in~\eqref{eq:Vrad}. A more complete analysis would involve performing a Weyl rescaling to go the Einstein frame, but the corrections this would introduce are suppressed by extra powers of $n(L)$ relative to the terms in~\eqref{eq:Veff} so can safely be ignored. We also note that the terms in the square brackets are proportional to the boundary conditions satisfied by $n(y)$, so vanish at the minimum. 

The effective potential can then be determined by solving the equations of motion for $n$ and $\Phi$ for different values of $L$ and substituting them into~\eqref{eq:Veff}. The bulk solutions for $\Phi$ and $n$ are determined by the EOM:
\es{eq:GWeom}{
    \Phi'' + \frac{4n'}{n} \Phi' = \frac{\D V}{\D \Phi}   \, ,
}
along with the Einstein equations~\eqref{eq:G00} and~\eqref{eq:G55}. To solve this system we write the warp factor as 
\begin{align}
    n(y) = n_0(y) + n_1(y) \, ,
\end{align}
where $n_0$ is given by the $\Phi = 0$ solution (see eq.~\eqref{eq:n_sol})
\begin{align}
    n_0 (y) &= \cosh (ky) -\sqrt{1+ (H/k)^2} \, \sinh(ky)\, ,
    \label{eq:warp_0}
\end{align}
and $n_1(y)$ captures the backreaction on the warp factor from $\Phi$, which must be included so that~\eqref{eq:Veff} captures all the contributions from the stabilizing sector~\cite{Lust:2025vyz}. Assuming this backreaction is small, however, we can solve the $\Phi$ EOM by setting $n = n_0$. Doing so allows for an analytic solution in terms of hypergeometric functions: 
\es{eq:phi_sol}{
    \Phi(y) &= c_+ \Phi_+(y) + c_- \Phi_-(y),
    \\
    \Phi_\pm(y) &= e^{a_\pm y} {}_2F_1 \left( 2, a_\pm ; a_\pm -1; - b e^{2y}  \right),
}
where:
\begin{align*}
    &a_\pm = 2 \pm \sqrt{4+\varepsilon} \, ,
    &&b =  \frac{1 - \sqrt{1+(H/k)^2}}{1+ \sqrt{1+(H/k)^2}} \, .
\end{align*}
After determining the solution for $\Phi$, it can be put back into the Einstein equations to determine $n_1(y)$:
\begin{align}
    n_1'' - k^2 n_1 + \frac{n_0 + n_1}{24 M_5^3} \left( 3 \Phi'^2 + \varepsilon \Phi^2 \right) = 0 \, .
    \label{eq:EOM_n1}
\end{align}
This equation has no analytic solution for general $\Phi$, so must be solved numerically for a given choice of parameters.

The main approximation we make in this procedure is using the unbackreacted warp factor, $n_0$, in eq.~\eqref{eq:GWeom} when solving for $\Phi$. A more complete solution would involve solving the coupled equations for $\Phi$ and $n$ numerically, but as this is a boundary-value problem the boundary conditions are difficult to implement. A common solution to this type of problem is to use a shooting method, varying the profiles $\Phi, n$ and their derivatives in the UV so that the profiles reach appropriate values in the IR. However, this is complicated by the fact that $\Phi$ and $n$ grow/fall exponentially, so small changes in the UV lead to exponentially large changes in the IR. Thus, for simplicity, we adopt the following approach: 
\begin{itemize}
    \item We set $n =n_0$ in eq.~\eqref{eq:GWeom}, which allows us to derive an analytic solution for $\Phi$, where we can use the boundary conditions~\eqref{eq:phi_bc} to fix the coefficients $c_\pm$ in eq.~\eqref{eq:phi_sol}.
    \item We then solve eq.~\eqref{eq:EOM_n1} for $n_1$ with the boundary condition $n_1(0) = n_1'(0) = 0$. This preserves the normalization condition $n(0) = 1$ and the UV boundary condition satisfied by $n_0$, which sets $H$, is unchanged.
    \item With these choices we then solve for $n_1$ and can adjust the detuning parameter $\tilde{V}$ (see eq.~\eqref{eq:bdy_potentials}) so that the boundary condition for $n$ is satisfied on the IR boundary. 
\end{itemize} 

\subsection{Numerical Solutions}

\begin{figure}[t!]
    \begin{tabular}{lc}
    \hspace{28mm} \textbf{Benchmark 1}  &   
     \hspace{12mm} \textbf{Benchmark 2}   \vspace{2mm} \\
    \includegraphics[scale=.8]{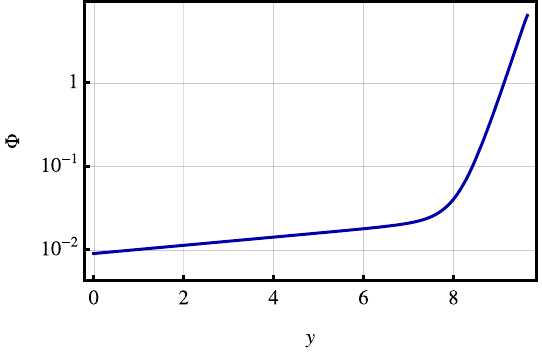}
    &
    \includegraphics[scale=.8]{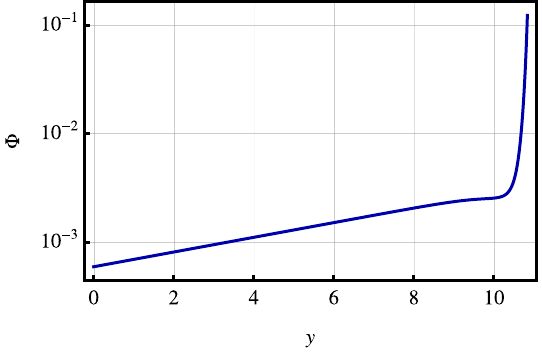}
    \\
    \includegraphics[scale=.8]{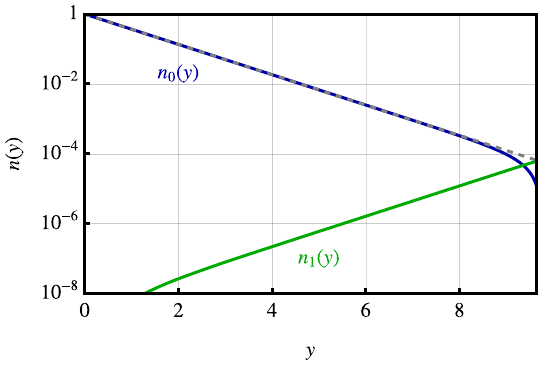}
    &
    \includegraphics[scale=.8]{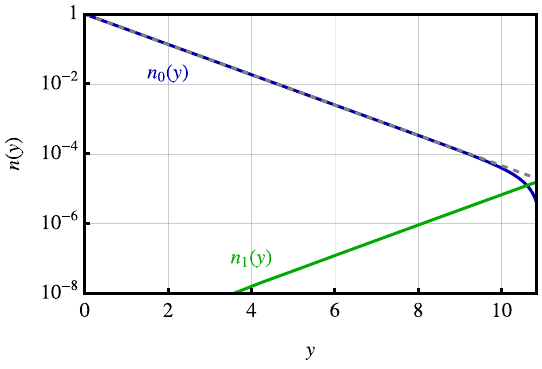}
    \\
    \end{tabular}
    \caption{Figures showing the Goldberger-Wise field profile (top) and warp factor (bottom) for the benchmark parameter points in eqs.~\eqref{eq:B1} and~\eqref{eq:B2}. Units are set so that $k=1$. The grey dashed lines in the bottom plots show the $H=0$ warp factor, which deviate from $n_0(y)$ at large $y$. We observe that, except in a region close to the IR boundary, $n_1$ represents a small correction to $n_0$.}
    \label{fig:profiles}
\end{figure}

The coefficients $c_\pm$ in~\eqref{eq:phi_sol} are determined by the boundary conditions~\eqref{eq:phi_bc}, which depend on the choice of boundary potentials. For concreteness, we take the boundary potentials to be given by
\begin{align}
    &v_{0} (\Phi) = k \lambda_{0} \left( \Phi - k^{3/2}f_{0} \right)^2  + \delta V_{\uv}\, ,
    &&v_{L} (\Phi) = k \lambda_{L} \left( \Phi - k^{3/2}f_{L} \right)^2 \, ,
\end{align}
where we include factors of $k$ to make the quartic coupling $\lambda_i$ and the VEV $f_i$ dimensionless. $\delta V_{\uv}$ is a constant tension which we adjust so that $v_0 (\Phi) = 0$ after solving for $\Phi$ -- this is not necessary, but ensures that $V_{\rm inf}$ directly controls the Hubble scale as in eq.~\eqref{eq:Vinf_sol}, without additional contributions from $v_0$. The model is completely determined by the parameters $\lambda_i, f_i, \varepsilon$, which determine the GW profile, the Hubble scale $H$ and the large-$N$ parameter (after choosing $M_4 = \mpl$). The numerical solutions are most easily found if the hierarchies between scales $H < k < \mpl$ are as small as possible, so we take $H=2 \times 10^{13}$ GeV, close to saturating the current bound from ACT/Planck~\cite{Planck:2018jri, ACT:2025tim}. We give the parametric dependence on $H$ where possible so our results can be extrapolated to smaller $H$.

For the other parameters we consider the benchmark points:
\begin{itemize}
    \item[]\textbf{Benchmark 1}
    \begin{align} \label{eq:B1}
        \epsilon = -0.44 , \, \lambda_{0} = \lambda_{L} = 1, \, f_0 = 8 \times 10^{-3}, \, f_L = 23 , \, N =10 , \, \alpha = 6.8 \times 10^{-9}, \, \beta = - 0.2   , \,
    \end{align}
    \item[]\textbf{Benchmark 2}
    \begin{align}\label{eq:B2}
        \epsilon = - 0.6 , \, \lambda_{0} = 0.8, \,  \lambda_{L} = 1, \, f_0 = 5\times 10^{-4}, \, f_L = 4.16 , \, N =3 , \, \alpha = 6.2 \times 10^{-10}, \, \beta = - 0.24 .
    \end{align}
\end{itemize}
In the first benchmark $M_5 = 8 \times 10^{17}$ GeV, $k=1.7 \times 10^{17}$ GeV to reproduce correct $\mpl$ in the $H\to 0$ limit, while for benchmark 2, $M_5 = 1.2 \times 10^{18}$ GeV and $k=5.7 \times 10^{17}$ GeV.

\begin{figure}[t!]
    \centering
    \includegraphics[scale=.75]{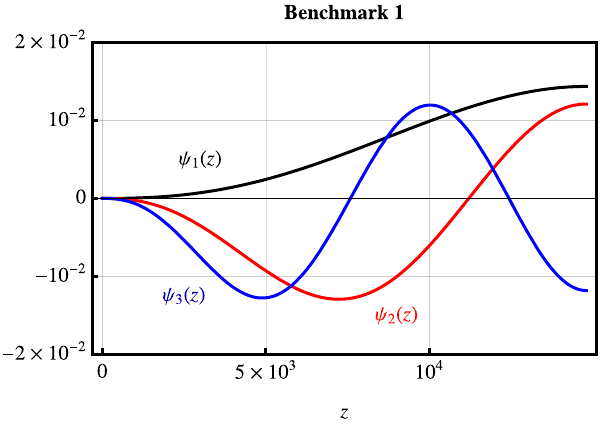}
    \includegraphics[scale=.75]{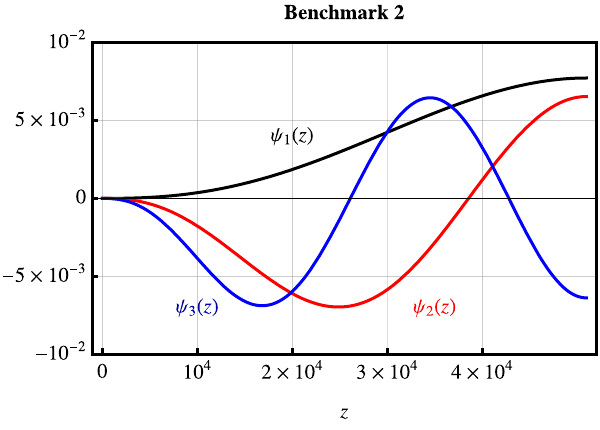}
    \caption{Plots showing the extra dimensional profiles for the first 3 KK modes for each of the benchmark points. The co-ordinate used is the homogeneous co-ordinate $z$, with the metric given in \eqref{eq:metric_z}. Units are such that $k=1$.}
    \label{fig:KKspectrum}
\end{figure}

In both cases we find that $\Lambda$ is stabilized at the Hubble scale, with $m_\varphi$ and the mass of the lightest KK graviton mode, $m_1$, also set by $H$. The first benchmark corresponds to a case where we are safely in the large-$N$ regime, but the NG signal is difficult to observe with current data as the couplings of the radion and KK gravitons are suppressed by $N$. However, these signatures would be observable in proposed 21-cm surveys~\cite{Bordin:2019tyb, Floss:2022grj}. For this reason we include the second benchmark, which has a smaller $N$. While this may be more sensitive to quantum gravity corrections, it leads to observable radion-mediated NG and more promising KK graviton-mediated NG, as shown below. 

We choose negative values for $\varepsilon$, so that $\Phi$ grows in the IR. This has a natural holographic interpretation as a marginally relevant coupling. The magnitude of $\varepsilon$ determines how quickly $\Phi$ grows, at least in the region where $y$ is small enough that $H$ can be ignored. Smaller values of $|\varepsilon|$ will lead to an IR boundary stabilized closer to the horizon. In the IR, when $n_0$ starts to deviate from the $H=0$ solution, $\Phi$ starts to grow exponentially. In each case $\Phi\sim \mathcal{O}(M_5)$ in the IR, which is necessary for the radion mass to be positive after including the tachyonic contribution from $H$ (see eq.~\eqref{eq:tachyon_mass}). The solutions for $\Phi, \, n_0$ and $n_1$ for each benchmark are shown in figure~\ref{fig:profiles}. Here we can see that $n_1 \ll n_0$ except for a small range of $y$ values close to the IR boundary, implying that the approximation we made of using $n_0$ in the equation of motion for $\Phi$ is good except in that range. As a check, we numerically solved for $\Phi$ using $n = n_0 + n_1$, with $n_1$ calculated using the analytic solution for $\Phi$ in eq.~\eqref{eq:phi_sol}, and found that this does not significantly alter our results. With the numerical solutions for $n$ and $\Phi$, we can also determine the KK graviton spectrum by solving the eigenvalue problem in eq.~\eqref{eq:eigen}. We show the wavefunctions, $\psi_l(z)$, of the lowest-lying KK modes for both the benchmarks in figure~\ref{fig:KKspectrum}, as a function of the co-ordinate $z$ defined in eq.~\eqref{eq:metric_z}.

\subsection{Spectrum of the Stabilized 5D Model}

For these benchmark points, the relevant parameters for the radion-mediated and KK graviton-mediated NG follow from eqs.~\eqref{eq:LsigmaKK} and~\eqref{eq:rad_couple_1}, and are given in table~\ref{tab:results}. These results are tree-level predictions, we comment on possible high-order corrections in appendix~\ref{app:higher}.
\begin{table}[h!]
\begin{center}
\begin{tabular}{|c|c|c|}
\hline
     & \textbf{Benchmark 1} &   \textbf{Benchmark 2}\\  
     \hline 
     $ \Lambda/H$   & 0.6   &    0.6    \\
     $ m_\varphi/H$     & 1.7   &    1.7   \\
     $ m_1/H$           & 1.8  &   1.8  \\
     $ \psi_1(z_\ir) \times 10^2$       & 1.4 &    0.8  \\
     $ F_\varphi/k$     & 56    &    9   \\
     $\lambda_{\varphi \sigma} $     &   -1.8   &   -2   \\
     $\langle \varphi \rangle /H $     &   31   &   5     \\
     \hline
\end{tabular}
\end{center}
\caption{Table summarizing the results obtained for each of the benchmark parameter points.}
\label{tab:results}
\end{table}

Our numerical results indicate the scaling
\begin{align}
    &\frac{m_\varphi}{\Lambda} \simeq \mathcal{O}(1) \, ,
    &&\frac{m_1}{\Lambda} \simeq \mathcal{O}(1) \, ,
\end{align}
which is expected from the $H= \Phi =0$ case where the KK wavefunctions can be solved exactly. Each of our parameter points also satisfy the bound $m_1 \geq 3H/2$, as expected from the $\Phi = 0$ solutions (see eq.~\eqref{eq:KK_wavefn_1}). We found that we required a strong backreaction from $\Phi$ in order to stabilize the radion with $H\neq 0$, with $\Phi$ reaching field values $\Phi, \Phi' \sim \mathcal{O} (M_5)$ in the IR for our benchmarks. This indicates that the scale invariance in the dual theory is badly broken, and $m_\varphi$ is set by $\Lambda$ -- without the suppression from approximate scale invariance that occurs in some models~\cite{Coradeschi:2013gda}. We also note that in each case $\langle \varphi \rangle > H$, in accordance with swampland arguments~\cite{Montero:2021otb, Mishra:2022fic}, meaning that radion oscillations around the minimum during inflation are negligible.

\section{Non-Gaussianity from the Radion and KK Gravitons}

\label{sec:NG}

In this section we calculate the NG signal for each of the benchmark points outlined in the previous section and compare to observational constraints from the CMB. In our scenario, the three-point function $Q = \zeta(\vec{k}_1) \zeta(\vec{k}_2) \zeta(\vec{k}_3)$ is suppressed by the soft breaking parameter $m_\sigma$, the curvaton mass, which we take to be $m_\sigma\ll H$ (see the discussion below eq.~\eqref{eq:f-decay}). Hence the leading NG signal in our model comes from the four-point function. 

\begin{figure}
    \centering
    \includegraphics[width=0.8\linewidth]{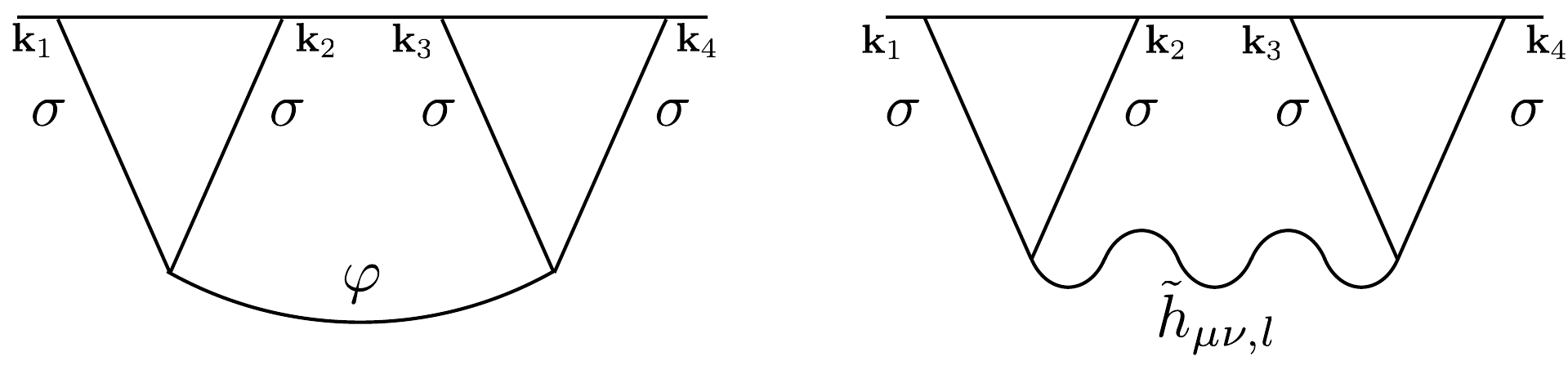}
    \caption{In-in diagrams showing the radion ($\varphi$)-mediated (left) and KK graviton ($\tilde{h}_{\mu\nu,l}$) mediated (right) trispectra. Time flows vertically upwards and the horizontal lines denote the end of inflation, parametrized by $t_0$. Both the radion and the KK graviton decay into curvaton ($\sigma$) fluctuations.}
    \label{fig:inin}
\end{figure}

We calculate the NG using the `in-in' formalism,  in which the expectation value of a generic operator $Q$, is given by, 
\es{eq:inin}{
    \langle Q\rangle = \langle \Omega  | U^\dagger Q_I(t_0) U |\Omega \rangle,
}
where $U = T\exp(-i\int_{-\infty(1-i\epsilon)}^{t_0} \D t \, \mathcal{H}_I^{\rm int}(t))$ is the time evolution operator written in terms of the interacting part of the Hamiltonian in the interaction picture, $\mathcal{H}^{\rm int}_I$. The relevant in-in diagrams mediating NG are shown in Fig.~\ref{fig:inin}, and the associated operator is $Q = \zeta(\vec{k}_1) \zeta(\vec{k}_2) \zeta(\vec{k}_3) \zeta(\vec{k}_4)$. 
In the scenario we are interested in the curvature perturbation $\zeta$ is dominantly sourced by curvaton fluctuations $\delta\sigma$,
\begin{align}
    \zeta = \frac{2}{3} \frac{\delta \sigma}{\sigma_i} \, ,
    \label{eq:zeta}
\end{align}
where $\sigma_i$ is the background value for $\sigma$. In eq.~\eqref{eq:zeta} we have assumed an $ m_\sigma^2 \sigma^2$ potential, $\fdec = 1$, and chosen a gauge where the scalar fluctuations of the spatial metric vanish.

\subsection{Radion-mediated Non-Gaussianity}

We first compute the scalar NG signal generated by radion exchange. We focus on the coupling of the radion to the kinetic term for $\sigma$ as this gives the dominant contributions, with additional couplings being suppressed by $m_\sigma$ (see discussion in sec.~\ref{subsec:couplings}). The bulk-to-boundary propagators describing the massless curvaton final states are given by
\es{}{
    G_\A(k,\eta) = {H^2\over 2k^3}(1-i\A k\eta)\exp(i\A k\eta),
}
where $\A = \pm$ denotes whether the bulk end of the propagator comes from time-ordering or anti time-ordering operator in eq.~\eqref{eq:inin}.\footnote{In this section, $k$ denotes the magnitude of three-momentum $\vec{k}$, and not the AdS curvature scale.} The quantized radion field can be written in terms of mode functions as,
\es{}{
    \delta\varphi(\vec{k},\eta) = f(k,\eta)^* a_{\vec{k}}^\dagger + f(k,\eta) a_{-\vec{k}},
}
with 
\es{}{
    f(k,\eta) = -{iH\sqrt{\pi} \over 2} \exp \left( \frac{i \pi}{2} \left( \frac12 +  i \mu_\varphi\right) \right)
    (-\eta)^{3/2}H^{(1)}_{i\mu_\varphi}(-k\eta),
}
and $\mu_\varphi = \sqrt{m_\varphi^2/H^2-9/4}$. In terms of these mode functions, the four bulk-to-bulk propagators for the radion can be written as,
\es{}{
    D_{-+}(k,\eta_1,\eta_2) &= f(k, \eta_1) f(k, \eta_2)^*,\\
    D_{+-}(k,\eta_1,\eta_2) &= f(k, \eta_1)^* f(k, \eta_2),\\
    D_{++}(k,\eta_1,\eta_2) &= D_{-+}(k,\eta_1,\eta_2)\theta(\eta_1-\eta_2) + D_{+-}(k,\eta_1,\eta_2)\theta(\eta_2-\eta_1),\\
    D_{--}(k,\eta_1,\eta_2) &= D_{+-}(k,\eta_1,\eta_2)\theta(\eta_1-\eta_2) + D_{-+}(k,\eta_1,\eta_2)\theta(\eta_2-\eta_1).
}
The leading interaction Lagrangian follows from~\eqref{eq:rad_couple_1}, 
\es{}{
    \int \D^4 x \left(- {\lambda_{\varphi \sigma}\over 2} \frac{\delta\varphi}{\langle \varphi\rangle} \right)  \sqrt{-g} g^{\mu\nu}\nabla_\mu\sigma\nabla_\nu\sigma.
}
Therefore, the four-point function can be computed by using the above interaction Lagrangian and the propagators for the curvaton and the radion. The result is:
\es{eq:trispec_radion}{
    &\langle \delta\sigma(\vec{k}_1) \delta\sigma(\vec{k}_2) \delta\sigma(\vec{k}_3) \delta\sigma(\vec{k}_4)\rangle'         \\ 
    &={\lambda_{\varphi \sigma}^2 \over H^4 \langle\varphi\rangle^2} \int_{-\infty}^0 \int_{-\infty}^0 {\D\eta_1 \over\eta_1^4} {\D \eta_2 \over \eta_2^4} 
    \left(-\eta_1^2 \partial_{\eta_1}G_\A(k_1,\eta_1)\partial_{\eta_1}G_\A(k_2,\eta_1) - \eta_1^2 (\vec{k}_1\cdot\vec{k}_2) G_\A(k_1,\eta_1) G_\A(k_2,\eta_1)\right)   \\
    &\times \left(-\eta_2^2 \partial_{\eta_2}G_\B(k_3,\eta_2)\partial_{\eta_2}G_\B(k_4,\eta_2) - \eta_2^2 (\vec{k}_3\cdot\vec{k}_4) G_\B(k_3,\eta_2) G_\B(k_4,\eta_2)\right)D_{\A\B}(k_I,\eta_1,\eta_2) + 2~{\rm perms.},
}
with $\vec{k}_I=\vec{k}_1+\vec{k}_2$. The two permutations include the $t$ and the $u$ channels. The above integral can be computed numerically.

To parametrize the NG mediated by the radion, we recall the definition of the local $\tau_{\rm NL}^{\rm loc}$ parameter describing the trispectrum~\cite{Smith:2015uia}, 
\es{}{
    \langle \zeta(\vec{k}_1) \zeta(\vec{k}_2) \zeta(\vec{k}_3) \zeta(\vec{k}_4)\rangle' = \tau_{\rm NL}^{\rm loc}P_\zeta(k_1)P_\zeta(k_3)P_\zeta(k_I) + 11~{\rm perms.} \, .
}
Here and above, the prime on the correlator denotes that the three-momentum-conserving delta functions have been taken out, i.e., $\langle \zeta(\vec{k}_1) \cdots \zeta(\vec{k}_n)\rangle = (2\pi)^3 \delta^3(\vec{k}_1+\cdots+\vec{k}_n) \langle \zeta(\vec{k}_1) \cdots \zeta(\vec{k}_n)\rangle'$. We fix $k_1=k_2=k_3=k_4=k$ and consider the explicitly written contribution in~\eqref{eq:trispec_radion} to illustrate the size of the trispectrum. In this case, the trispectrum depends only on $k$ and $k_I=|\vec{k}_1+\vec{k}_2|$, and we {\it define} its strength via,
\es{}{
    \langle \zeta(\vec{k}_1) \zeta(\vec{k}_2) \zeta(\vec{k}_3) \zeta(\vec{k}_4)\rangle'\bigg\rvert_{|\vec{k}_i|=k, \, k_I \lesssim k} = 
    \tau_{\rm NL}^{\rm radion} \, \left({k_I \over k}\right)^3 P_\zeta(k)^2 P_\zeta(k_I) \, .
}
Using $P_\zeta(k) =\langle\zeta(\vec{k})\zeta(-\vec{k})\rangle' = 2H^2/(9\sigma_i^2 k^3)$, we get 
\es{}{
    \tau_{\rm NL}^{\rm radion} =  {18 \lambda_{\varphi \sigma}^2 \sigma
    _i^2 \over \langle\varphi\rangle^2} T_{\rm radion}(k_I/k, \mu_\varphi),
}
where $T_{\rm radion}$ is a dimensionless function of the momentum ratio $k_I/k$ and the radion mass, which we compute numerically. As per the cosmological collider physics mechanism, $T_{\rm radion}$ describes the on-shell propagation of the radion before its decay into curvaton fluctuations. Accordingly, the temporal oscillations of the radion translates into oscillations in $k_I/k$~\cite{Arkani-Hamed:2015bza}, as shown explicitly in Fig.~\ref{fig:trispec}. Parametrically, the prefactor in $\tau_{\rm NL}^{\rm radion}$ can be expressed as,
\es{eq:tau_rad}{
    {\lambda_{\varphi \sigma}^2 \sigma
    _i^2 \over \langle\varphi\rangle^2} \sim {\lambda_{\varphi \sigma}^2 \sigma
    _i^2 \over N^2 \Lambda^2},
}
which illustrates the scaling with $N$.

\subsection{KK Graviton-mediated Non-Gaussianity}

To evaluate the spin-2 NG signal mediated by the exchange of KK gravitons, it is useful to decompose the $l$-th KK graviton, with five degrees of freedom, into helicity eigenstates $\tilde{h}_{\mu\nu,l} = \sum_{\lambda=-2}^{2}\tilde{h}_{\mu\nu,l}^{(\lambda)}$, where $\tilde{h}_{\mu\nu,l}$ satisfies eq.~\eqref{eq:htildemunu}. While all the helicity modes would mediate NG, here we focus on the helicity $\pm 2$ modes, which are unique to a massive spin-2 particle. The helicity $\pm 2$ components can be isolated by writing $\tilde{h}_{ij,l}^{(\pm 2)} = h_l^{(\pm 2)}\epsilon_{ij}^{(\pm 2)} + \cdots$, where $\cdots$ includes helicity $0$ and $\pm 1$ states. For a mode propagating along the $\hat{z}$ direction, the polarization tensor is given by
\es{}{
    \epsilon_{ij}^{(\pm 2)}(\hat{z}) = \begin{pmatrix}
        1 & \pm i & 0 \\ \pm i & -1 & 0 \\ 0 & 0 & 0
    \end{pmatrix}.
}
The mode functions are given by
\es{}{
    h_{l}^{\pm 2}(k,\eta) = {\sqrt{\pi k} \over 2H} 
    \exp \left( \frac{i \pi}{2} \left( \frac12 +  i \mu \right) \right)
    %\exp(i\pi/4) \exp(-\pi \mu/2)
    (-k\eta)^{-1/2}H^{(1)}_{i\mu}(-k\eta)
}
with $\mu = \sqrt{m^2/H^2-9/4}$, and conjugate of $h_{l}^{\pm 2}(k,\eta)$. In terms of these mode functions, the four bulk-to-bulk propagators can be written as,
\es{}{
    \tilde{D}_{-+}(k,\eta_1,\eta_2) &= h_{l}^{\pm 2}(k, \eta_1) h_{l}^{\pm 2}(k, \eta_2)^*,\\
    \tilde{D}_{+-}(k,\eta_1,\eta_2) &= h_{l}^{\pm 2}(k, \eta_1)^* h_{l}^{\pm 2}(k, \eta_2),\\
    \tilde{D}_{++}(k,\eta_1,\eta_2) &= \tilde{D}_{-+}(k,\eta_1,\eta_2)\theta(\eta_1-\eta_2) + \tilde{D}_{+-}(k,\eta_1,\eta_2)\theta(\eta_2-\eta_1),\\
    \tilde{D}_{--}(k,\eta_1,\eta_2) &= \tilde{D}_{+-}(k,\eta_1,\eta_2)\theta(\eta_1-\eta_2) + \tilde{D}_{-+}(k,\eta_1,\eta_2)\theta(\eta_2-\eta_1).
}
The interaction Lagrangian follows from~\eqref{eq:LsigmaKK} 
\es{}{
    -{1\over 2 M_4} \chi_l(L) \int \D^4 x \sqrt{-\bar{g}} \nabla^\mu\delta\sigma\nabla^\nu\delta\sigma \tilde{h}_{\mu\nu,l} \supset -{1\over 2 M_4} \chi_l(L) \int {\D \eta}\D^3\vec{x}\partial_i\delta\sigma \partial_j\delta\sigma h_l^{(\pm 2)}\epsilon_{ij}^{(\pm 2)} +\cdots.
}
Therefore, for each helicity we have,
\es{}{
    &\langle \delta\sigma(\vec{k}_1) \delta\sigma(\vec{k}_2) \delta\sigma(\vec{k}_3) \delta\sigma(\vec{k}_4)\rangle' \\ &={\chi_l(L)^2 \over M_4^2} P(\vec{k}_1,\vec{k}_3)\int_{-\infty}^0 \D\eta_1 \D \eta_2 G_\A(k_1,\eta_1)G_\A(k_2,\eta_1)G_\B(k_3,\eta_2)G_\B(k_4,\eta_2)\tilde{D}_{\A\B}(k_I,\eta_1,\eta_2) + 2~{\rm perms.},
}
where $P(\vec{k}_1,\vec{k}_3) = k_{1i}k_{2j}k_{3m}k_{4n} \epsilon_{ij}^{(\pm 2)}(\hat{k}_I) \epsilon_{mn}^{(\pm 2)}(-\hat{k}_I)=k_1^2 k_3^2\sin^2(\theta_1)\sin^2(\theta_3)\exp(\mp 2i\phi_3)$ with $\vec{k}_I =\vec{k}_1+\vec{k}_2$, $\cos\theta_1 = \hat{k}_1\cdot \hat{k}_I$, $\cos\theta_3 = \hat{k}_3\cdot \hat{k}_I$, and $\phi_3$ is the dihedral angle between the two planes containing $\vec{k}_{1,2}$ and $\vec{k}_{3,4}$. The contribution shown explicitly above is from the $s$-channel and the two permutations include the $t$ and $u$ channels.

\begin{figure}[t]
    \begin{tabular}{lc}
    \hspace{30mm} \textbf{Benchmark 1}  &   
     \hspace{12mm} \textbf{Benchmark 2}   \vspace{2mm} \\
    \hspace{1mm}
    \includegraphics[scale=.7 ]{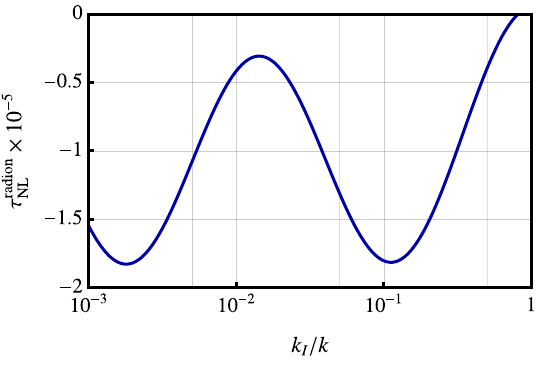}
    &\hspace{3mm}
    \includegraphics[scale=.7]{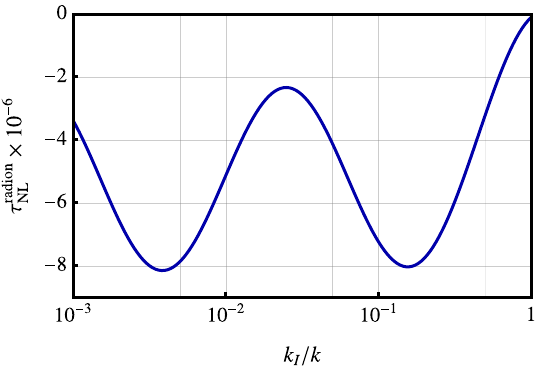}
    \\
    \includegraphics[scale=.73]{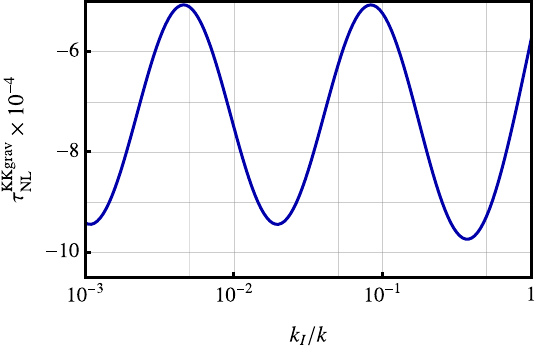}
    &
    \includegraphics[scale=.75]{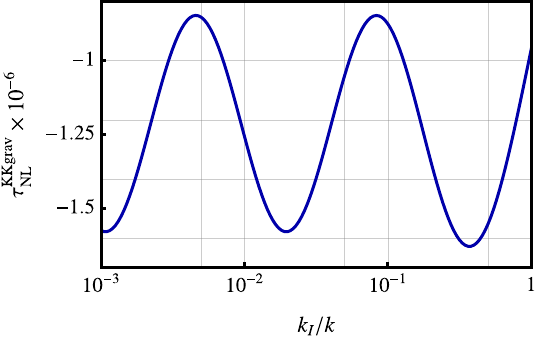}
    \\
    \end{tabular}
    \caption{Trispectrum mediated by the radion (top) and the KK graviton (bottom). The oscillations capture the on-shell production of the radion and the KK graviton during inflation.}
    \label{fig:trispec}
\end{figure}

Following the same procedure as above, we can define a KK graviton-mediated trispectrum,
\es{}{
    \langle \zeta(\vec{k}_1) \zeta(\vec{k}_2) \zeta(\vec{k}_3) \zeta(\vec{k}_4)\rangle\bigg\rvert_{|\vec{k}_i|=k, \, k_I \lesssim k} = 
    \tau_{\rm NL}^{\rm KK grav} \, \left({k_I \over k}\right)^3 P_\zeta(k)^2 P_\zeta(k_I) \sin^2\theta_1 \sin^2\theta_3 \cos(2\phi_3) .
}
Taking into account the $s$-channel contribution,
\es{}{
    \tau_{\rm NL}^{\rm KK grav} = {18 \chi_l(L)^2 \sigma_i^2 \over M_4^2} T_{\rm KK grav}(k_I/k,\mu).
}
Similar to the radion, $T_{\rm KK grav}$ oscillates as a function of $k_I/k$ (Fig.~\ref{fig:trispec}), depicting the on-shell propagation of the graviton following its production. The prefactor can be expressed as
\es{eq:tau_grav}{
    {\chi_l(L)^2 \sigma_i^2 \over  M_4^2} \sim {\sigma_i^2 \over N^2 \Lambda^2}\, ,
}
where we have used~\eqref{eq:normalisation}.

\subsection{Results and Future Prospects}

We now summarize the results of the numerical computation of the trispectrum shown in Fig.~\ref{fig:trispec} and compare that with expectations based on the 4D dual CFT. First, we focus on $\tau_{\rm NL}^{\rm radion}$. Based on eq.~\eqref{eq:tau_rad}, and taking $\lambda_{\varphi \sigma}\sim 2$, $\Lambda\sim H$, and $\sigma_i \sim 2\times 10^3 H$ to reproduce the correct power spectrum (eq.~\eqref{eq:ps}), we expect $\tau_{\rm NL}^{\rm radion} \sim 10^7/N^2$. This matches at an order-of-magnitude level the results seen in Fig.~\ref{fig:trispec}, where Benchmark 1 and 2 correspond to $N=10$ and $N=3$, respectively. For both these two benchmarks, the radion mass was given by $m_\varphi \approx 1.7 H$. Therefore, the `Boltzmann suppression', while present, is not severe: $\exp(-\pi(m_\varphi^2/H^2-9/4)^{1/2}) \sim 0.1$. For $N=3$, these values are comparable to the observational bound derived in~\cite{Philcox:2025wts} for a similarly massive spin-0 particle: $\tau_{\rm NL}^{\rm heavy} = (-5.8\pm 7.9)\times 10^6$. For a more detailed comparison, a full evaluation of the radion-mediated trispectrum, beyond the specific kinematic configuration considered here, is required. This will be addressed in a separate work. For $N=10$, the NG signal is an order of magnitude below the CMB bounds presented in~\cite{Philcox:2025wts}, but future observations, especially involving 21-cm surveys~\cite{Bordin:2019tyb, Floss:2022grj}, would provide powerful sensitivity.

The trispectrum mediated by the KK graviton, has an overall scaling which goes as $\sigma_i^2 /(N \Lambda)^2$ (see eq.~\eqref{eq:tau_grav}). This has exactly the same parametric dependence on $\sigma_i, \Lambda$, and $N$, as the radion-mediated trispectrum. This is expected from the 4D arguments in sec.~\ref{subsec:couplings} since both the radion and the KK gravitons are dual to the glueballs of the (broken) CFT. Based on these arguments, we estimate $\tau_{\rm NL}^{\rm KKgrav} \sim 4\times 10^6/N^2$ for $\Lambda\sim H$. This also matches at an order-of-magnitude level with the results in Fig.~\ref{fig:trispec}. For $N=3$, the trispectrum is somewhat smaller than the bounds obtained in~\cite{Philcox:2025wts} for a similarly massive spin-2 particle: $\tau_{\rm NL}^{\rm heavy} = (0.8\pm 2.9)\times 10^7$. However, we recall that we have only computed the contribution in specific kinematic configurations and considered only the helicity $\pm 2$ component of the KK graviton. A full computation, including the other helicity modes, would be relevant in making a concrete connection with the observational bounds. This will also be addressed in a separate work. For $N=10$, future 21-cm surveys will again be especially relevant in constraining this signature.

In the above, we have described how to stabilize an extra dimension such that both the radion and the lightest KK graviton have a mass of order the Hubble scale. However, extra dimensions can have sizes much smaller sizes such that both $m_\varphi, m_1 \gg H$. In such a scenario, the oscillatory signatures shown in Fig.~\ref{fig:trispec} would be exponentially suppressed as $\exp(-\pi m/H)$. Several mechanisms have been proposed that can overcome this sort of suppression in the context of standard 4D theories. Examples include `chemical potential'-like derivative couplings of the inflaton field with currents consisting of massive fields~\cite{Chen:2018xck, Hook:2019zxa, Bodas:2020yho, Wang:2020ioa, Tong:2022cdz}, and features on the inflationary potential~\cite{Chen:2022vzh, Pajer:2024ckd}. It would be interesting to explore whether these mechanisms could also be operational in the extra dimensional scenarios discussed in this work.

\section{Conclusions}
\label{sec:conclusions}

Cosmological observations offer a way to probe energy scales far above those accessible to terrestrial experiments. Particularly interesting in this regard are primordial non-Gaussianities mediated by states with masses of order $H$ during inflation. Given $H$ could be as large as $10^{13}$ GeV, this `cosmological collider' could extend our experimental reach by many orders of magnitude. This opportunity has led to a large research effort devoted to understanding the prospects for seeing new physics in these observations. An intriguing possibility is that new states from the presence of extra dimensions could lead to such a signal. Despite being well motivated from UV models such as string theory, studies of the cosmological collider signal from extra dimensions have focused on flat extra dimensions and have not fully accounted for the effects of stabilization.

In this work we studied the cosmological collider signal from warped extra dimensions. Scalar moduli and KK graviton modes are two generic predictions of extra dimensions, as they correspond to gravitational degrees of freedom in the higher dimensional theory. When the scale, $\Lambda$, of the extra dimension is $\Lambda \sim H$, both the radion modulus and KK states have masses of $\mathcal{O}(H)$ and can mediate a large NG signal. When the extra dimension is warped, the relevant couplings of the radion and KK modes to IR states are suppressed by $\Lambda$ rather than $\mpl$. This leads to an enhanced signal compared to flat extra dimensions if the curvature perturbation is generated by a field on the IR boundary. We explicitly showed this by implementing a mechanism to stabilize the extra dimension, which allowed us to calculate the radion mass, KK mass, and their couplings in a complete, stabilized 5D model. 

This setup thus leads to a spin-0 cosmological collider signal generated by the radion as well as a spin-2 signal generated by KK graviton modes. In both cases, the couplings which generate the signal are suppressed by factors of the large-$N$ parameter of the RS model, and by the Boltzmann factor $e^{-\pi m/H}$ characteristic of the cosmological collider signal. The signal is thus largest when $N$ is made small (while retaining theoretical control) and the masses of the mediating states are not too much larger than $H$. We found benchmark parameter points for $N = 3$ where the spin-0 non-Gaussianity is large enough to be constrained by current CMB data. The spin-2 signal in this case, and both the spin-0 and spin-2 signal for larger $N$, evade current bounds but may be observable in future surveys. This highlights the exciting prospect that current and upcoming cosmological observations may see a sign of high-scale extra dimensions.

\section*{Acknowledgments}

We thank Lisa Randall for collaborating on the early stages of this project. We also thank Arushi Bodas, Priyesh Chakraborty, Majid Ekhterachian, Matthew Lewandowski, Rashmish Mishra, Lisa Randall, Raman Sundrum, and Zhong-Zhi Xianyu for useful discussions. MN is supported in part by NSF Award PHY-2310717, and would like to acknowledge GRASP Initiative funding provided by Harvard University. This research was supported in part by grant NSF PHY-2309135 to the Kavli Institute for Theoretical Physics (KITP).

\appendix

\section{Higher Order Corrections}

\label{app:higher}

In this appendix, we discuss loop corrections to the masses of the radion and curvaton and show that they are generically suppressed, justifying our approximation of working to tree-level. Corrections to the radion mass come from fields $\chi$ on the IR boundary, which interact with the radion through the terms (after integrating over $y$)
\begin{align}
    \frac{\mathcal{L}_{\chi \varphi}}{\sqrt{-g}} = \frac{1}{2}\left( \frac{m_\chi}{\langle \varphi \rangle} \right)^2 \chi^2 \varphi^2  \, .
\end{align}
Expanding $\varphi$ around its VEV leads to a cubic interaction vertex $\delta\varphi \chi^2$, which generates a correction to the radion mass which goes like~\cite{Chen:2016hrz}
\begin{align}
    \delta m_\varphi^2 &= \frac{3 H^4}{8 \pi^2 \langle \varphi \rangle^2} \, ,
\end{align}
where we have assumed $m_\chi \lesssim H \lesssim \Lambda$ for IR-localized fields. The proportional shift in $m_\varphi$ is then
\begin{align}
    \frac{\delta m_\varphi^2}{m_\varphi^2} &= \frac{3 H^4}{8 \pi^2 \langle \varphi \rangle^2 \Lambda^2} \, .
\end{align}
The shift in the mass is therefore negligible, being suppressed by $1/N^2$, loop factors and~$\langle \varphi \rangle / H$. Similarly, corrections to the curvaton mass from radion loops are controlled by the coupling $\lambda = \left(m_\sigma /\langle \varphi \rangle \right)^2$, which is suppressed both as $\langle \varphi \rangle \sim N \Lambda$ and as we are working in the regime $m_\sigma \ll \Lambda$.

\bibliographystyle{utphys}
\bibliography{refs}

\providecommand{\href}[2]{#2}\begingroup\raggedright\begin{thebibliography}{100}

\bibitem{Arkani-Hamed:1998jmv}
N.~Arkani-Hamed, S.~Dimopoulos, and G.~R. Dvali, ``{The Hierarchy problem and
  new dimensions at a millimeter},''
  \href{http://dx.doi.org/10.1016/S0370-2693(98)00466-3}{{\em Phys. Lett. B}
  {\bfseries 429} (1998) 263--272},
  \href{http://arxiv.org/abs/hep-ph/9803315}{{\ttfamily arXiv:hep-ph/9803315}}.

\bibitem{Antoniadis:1998ig}
I.~Antoniadis, N.~Arkani-Hamed, S.~Dimopoulos, and G.~R. Dvali, ``{New
  dimensions at a millimeter to a Fermi and superstrings at a TeV},''
  \href{http://dx.doi.org/10.1016/S0370-2693(98)00860-0}{{\em Phys. Lett. B}
  {\bfseries 436} (1998) 257--263},
  \href{http://arxiv.org/abs/hep-ph/9804398}{{\ttfamily arXiv:hep-ph/9804398}}.

\bibitem{Arkani-Hamed:1998sfv}
N.~Arkani-Hamed, S.~Dimopoulos, and G.~R. Dvali, ``{Phenomenology, astrophysics
  and cosmology of theories with submillimeter dimensions and TeV scale quantum
  gravity},'' \href{http://dx.doi.org/10.1103/PhysRevD.59.086004}{{\em Phys.
  Rev. D} {\bfseries 59} (1999) 086004},
  \href{http://arxiv.org/abs/hep-ph/9807344}{{\ttfamily arXiv:hep-ph/9807344}}.

\bibitem{Randall:1999ee}
L.~Randall and R.~Sundrum, ``{A Large mass hierarchy from a small extra
  dimension},'' \href{http://dx.doi.org/10.1103/PhysRevLett.83.3370}{{\em Phys.
  Rev. Lett.} {\bfseries 83} (1999) 3370--3373},
  \href{http://arxiv.org/abs/hep-ph/9905221}{{\ttfamily arXiv:hep-ph/9905221}}.

\bibitem{Randall:1999vf}
L.~Randall and R.~Sundrum, ``{An Alternative to compactification},''
  \href{http://dx.doi.org/10.1103/PhysRevLett.83.4690}{{\em Phys. Rev. Lett.}
  {\bfseries 83} (1999) 4690--4693},
  \href{http://arxiv.org/abs/hep-th/9906064}{{\ttfamily arXiv:hep-th/9906064}}.

\bibitem{Grossman:1999ra}
Y.~Grossman and M.~Neubert, ``{Neutrino masses and mixings in nonfactorizable
  geometry},'' \href{http://dx.doi.org/10.1016/S0370-2693(00)00054-X}{{\em
  Phys. Lett. B} {\bfseries 474} (2000) 361--371},
  \href{http://arxiv.org/abs/hep-ph/9912408}{{\ttfamily arXiv:hep-ph/9912408}}.

\bibitem{Gherghetta:2000qt}
T.~Gherghetta and A.~Pomarol, ``{Bulk fields and supersymmetry in a slice of
  AdS},'' \href{http://dx.doi.org/10.1016/S0550-3213(00)00392-8}{{\em Nucl.
  Phys. B} {\bfseries 586} (2000) 141--162},
  \href{http://arxiv.org/abs/hep-ph/0003129}{{\ttfamily arXiv:hep-ph/0003129}}.

\bibitem{Huber:2000ie}
S.~J. Huber and Q.~Shafi, ``{Fermion masses, mixings and proton decay in a
  Randall-Sundrum model},''
  \href{http://dx.doi.org/10.1016/S0370-2693(00)01399-X}{{\em Phys. Lett. B}
  {\bfseries 498} (2001) 256--262},
  \href{http://arxiv.org/abs/hep-ph/0010195}{{\ttfamily arXiv:hep-ph/0010195}}.

\bibitem{CidVidal:2018eel}
X.~Cid~Vidal {\em et~al.}, ``{Report from Working Group 3}: {Beyond the
  Standard Model physics at the HL-LHC and HE-LHC},''
  \href{http://dx.doi.org/10.23731/CYRM-2019-007.585}{{\em CERN Yellow Rep.
  Monogr.} {\bfseries 7} (2019) 585--865},
  \href{http://arxiv.org/abs/1812.07831}{{\ttfamily arXiv:1812.07831
  [hep-ph]}}.

\bibitem{ATLAS:2024fdw}
{\bfseries ATLAS} Collaboration, G.~Aad {\em et~al.}, ``{Exploration at the
  high-energy frontier: ATLAS Run~2 searches investigating the exotic jungle
  beyond the Standard Model},''
  \href{http://dx.doi.org/10.1016/j.physrep.2024.10.001}{{\em Phys. Rept.}
  {\bfseries 1116} (2025) 301--385},
  \href{http://arxiv.org/abs/2403.09292}{{\ttfamily arXiv:2403.09292
  [hep-ex]}}.

\bibitem{CMS:2024nht}
{\bfseries CMS} Collaboration, A.~Hayrapetyan {\em et~al.}, ``{Search for new
  physics in high-mass diphoton events from proton-proton collisions at $
  \sqrt{\textrm{s}} $ = 13 TeV},''
  \href{http://dx.doi.org/10.1007/JHEP08(2024)215}{{\em JHEP} {\bfseries 08}
  (2024) 215}, \href{http://arxiv.org/abs/2405.09320}{{\ttfamily
  arXiv:2405.09320 [hep-ex]}}.

\bibitem{Giudice:2017fmj}
G.~F. Giudice, Y.~Kats, M.~McCullough, R.~Torre, and A.~Urbano,
  ``{Clockwork/linear dilaton: structure and phenomenology},''
  \href{http://dx.doi.org/10.1007/JHEP06(2018)009}{{\em JHEP} {\bfseries 06}
  (2018) 009}, \href{http://arxiv.org/abs/1711.08437}{{\ttfamily
  arXiv:1711.08437 [hep-ph]}}.

\bibitem{Murata:2014nra}
J.~Murata and S.~Tanaka, ``{A review of short-range gravity experiments in the
  LHC era},'' \href{http://dx.doi.org/10.1088/0264-9381/32/3/033001}{{\em
  Class. Quant. Grav.} {\bfseries 32} no.~3, (2015) 033001},
  \href{http://arxiv.org/abs/1408.3588}{{\ttfamily arXiv:1408.3588 [hep-ex]}}.

\bibitem{Tan:2016vwu}
W.-H. Tan, S.-Q. Yang, C.-G. Shao, J.~Li, A.-B. Du, B.-F. Zhan, Q.-L. Wang,
  P.-S. Luo, L.-C. Tu, and J.~Luo, ``{New Test of the Gravitational
  Inverse-Square Law at the Submillimeter Range with Dual Modulation and
  Compensation},'' \href{http://dx.doi.org/10.1103/PhysRevLett.116.131101}{{\em
  Phys. Rev. Lett.} {\bfseries 116} no.~13, (2016) 131101}.

\bibitem{Lee:2020zjt}
J.~G. Lee, E.~G. Adelberger, T.~S. Cook, S.~M. Fleischer, and B.~R. Heckel,
  ``{New Test of the Gravitational $1/r^2$ Law at Separations down to 52
  $\mu$m},'' \href{http://dx.doi.org/10.1103/PhysRevLett.124.101101}{{\em Phys.
  Rev. Lett.} {\bfseries 124} no.~10, (2020) 101101},
  \href{http://arxiv.org/abs/2002.11761}{{\ttfamily arXiv:2002.11761
  [hep-ex]}}.

\bibitem{Chen:2010xka}
X.~Chen, ``{Primordial Non-Gaussianities from Inflation Models},''
  \href{http://dx.doi.org/10.1155/2010/638979}{{\em Adv. Astron.} {\bfseries
  2010} (2010) 638979}, \href{http://arxiv.org/abs/1002.1416}{{\ttfamily
  arXiv:1002.1416 [astro-ph.CO]}}.

\bibitem{Wang:2013zva}
Y.~Wang, ``{Inflation, Cosmic Perturbations and Non-Gaussianities},''
  \href{http://dx.doi.org/10.1088/0253-6102/62/1/19}{{\em Commun. Theor. Phys.}
  {\bfseries 62} (2014) 109--166},
  \href{http://arxiv.org/abs/1303.1523}{{\ttfamily arXiv:1303.1523 [hep-th]}}.

\bibitem{Chen:2009zp}
X.~Chen and Y.~Wang, ``{Quasi-Single Field Inflation and Non-Gaussianities},''
  \href{http://dx.doi.org/10.1088/1475-7516/2010/04/027}{{\em JCAP} {\bfseries
  04} (2010) 027}, \href{http://arxiv.org/abs/0911.3380}{{\ttfamily
  arXiv:0911.3380 [hep-th]}}.

\bibitem{Baumann:2011nk}
D.~Baumann and D.~Green, ``{Signatures of Supersymmetry from the Early
  Universe},'' \href{http://dx.doi.org/10.1103/PhysRevD.85.103520}{{\em Phys.
  Rev. D} {\bfseries 85} (2012) 103520},
  \href{http://arxiv.org/abs/1109.0292}{{\ttfamily arXiv:1109.0292 [hep-th]}}.

\bibitem{Noumi:2012vr}
T.~Noumi, M.~Yamaguchi, and D.~Yokoyama, ``{Effective field theory approach to
  quasi-single field inflation and effects of heavy fields},''
  \href{http://dx.doi.org/10.1007/JHEP06(2013)051}{{\em JHEP} {\bfseries 06}
  (2013) 051}, \href{http://arxiv.org/abs/1211.1624}{{\ttfamily arXiv:1211.1624
  [hep-th]}}.

\bibitem{Chen:2012ge}
X.~Chen and Y.~Wang, ``{Quasi-Single Field Inflation with Large Mass},''
  \href{http://dx.doi.org/10.1088/1475-7516/2012/09/021}{{\em JCAP} {\bfseries
  09} (2012) 021}, \href{http://arxiv.org/abs/1205.0160}{{\ttfamily
  arXiv:1205.0160 [hep-th]}}.

\bibitem{Arkani-Hamed:2015bza}
N.~Arkani-Hamed and J.~Maldacena, ``{Cosmological Collider Physics},''
  \href{http://arxiv.org/abs/1503.08043}{{\ttfamily arXiv:1503.08043
  [hep-th]}}.

\bibitem{Planck:2018jri}
{\bfseries Planck} Collaboration, Y.~Akrami {\em et~al.}, ``{Planck 2018
  results. X. Constraints on inflation},''
  \href{http://dx.doi.org/10.1051/0004-6361/201833887}{{\em Astron. Astrophys.}
  {\bfseries 641} (2020) A10},
  \href{http://arxiv.org/abs/1807.06211}{{\ttfamily arXiv:1807.06211
  [astro-ph.CO]}}.

\bibitem{ACT:2025tim}
{\bfseries ACT} Collaboration, E.~Calabrese {\em et~al.}, ``{The Atacama
  Cosmology Telescope: DR6 Constraints on Extended Cosmological Models},''
  \href{http://arxiv.org/abs/2503.14454}{{\ttfamily arXiv:2503.14454
  [astro-ph.CO]}}.

\bibitem{Baumann:2011su}
D.~Baumann and D.~Green, ``{Equilateral Non-Gaussianity and New Physics on the
  Horizon},'' \href{http://dx.doi.org/10.1088/1475-7516/2011/09/014}{{\em JCAP}
  {\bfseries 09} (2011) 014}, \href{http://arxiv.org/abs/1102.5343}{{\ttfamily
  arXiv:1102.5343 [hep-th]}}.

\bibitem{Assassi:2013gxa}
V.~Assassi, D.~Baumann, D.~Green, and L.~McAllister, ``{Planck-Suppressed
  Operators},'' \href{http://dx.doi.org/10.1088/1475-7516/2014/01/033}{{\em
  JCAP} {\bfseries 01} (2014) 033},
  \href{http://arxiv.org/abs/1304.5226}{{\ttfamily arXiv:1304.5226 [hep-th]}}.

\bibitem{Craig:2014rta}
N.~Craig and D.~Green, ``{Testing Split Supersymmetry with Inflation},''
  \href{http://dx.doi.org/10.1007/JHEP07(2014)102}{{\em JHEP} {\bfseries 07}
  (2014) 102}, \href{http://arxiv.org/abs/1403.7193}{{\ttfamily arXiv:1403.7193
  [hep-ph]}}.

\bibitem{Dimastrogiovanni:2015pla}
E.~Dimastrogiovanni, M.~Fasiello, and M.~Kamionkowski, ``{Imprints of Massive
  Primordial Fields on Large-Scale Structure},''
  \href{http://dx.doi.org/10.1088/1475-7516/2016/02/017}{{\em JCAP} {\bfseries
  02} (2016) 017}, \href{http://arxiv.org/abs/1504.05993}{{\ttfamily
  arXiv:1504.05993 [astro-ph.CO]}}.

\bibitem{Lee:2016vti}
H.~Lee, D.~Baumann, and G.~L. Pimentel, ``{Non-Gaussianity as a Particle
  Detector},'' \href{http://dx.doi.org/10.1007/JHEP12(2016)040}{{\em JHEP}
  {\bfseries 12} (2016) 040}, \href{http://arxiv.org/abs/1607.03735}{{\ttfamily
  arXiv:1607.03735 [hep-th]}}.

\bibitem{Meerburg:2016zdz}
P.~D. Meerburg, M.~M\"unchmeyer, J.~B. Mu\~noz, and X.~Chen, ``{Prospects for
  Cosmological Collider Physics},''
  \href{http://dx.doi.org/10.1088/1475-7516/2017/03/050}{{\em JCAP} {\bfseries
  03} (2017) 050}, \href{http://arxiv.org/abs/1610.06559}{{\ttfamily
  arXiv:1610.06559 [astro-ph.CO]}}.

\bibitem{Chen:2016uwp}
X.~Chen, Y.~Wang, and Z.-Z. Xianyu, ``{Standard Model Background of the
  Cosmological Collider},''
  \href{http://dx.doi.org/10.1103/PhysRevLett.118.261302}{{\em Phys. Rev.
  Lett.} {\bfseries 118} no.~26, (2017) 261302},
  \href{http://arxiv.org/abs/1610.06597}{{\ttfamily arXiv:1610.06597
  [hep-th]}}.

\bibitem{Chen:2016nrs}
X.~Chen, Y.~Wang, and Z.-Z. Xianyu, ``{Loop Corrections to Standard Model
  Fields in Inflation},'' \href{http://dx.doi.org/10.1007/JHEP08(2016)051}{{\em
  JHEP} {\bfseries 08} (2016) 051},
  \href{http://arxiv.org/abs/1604.07841}{{\ttfamily arXiv:1604.07841
  [hep-th]}}.

\bibitem{Chen:2016hrz}
X.~Chen, Y.~Wang, and Z.-Z. Xianyu, ``{Standard Model Mass Spectrum in
  Inflationary Universe},''
  \href{http://dx.doi.org/10.1007/JHEP04(2017)058}{{\em JHEP} {\bfseries 04}
  (2017) 058}, \href{http://arxiv.org/abs/1612.08122}{{\ttfamily
  arXiv:1612.08122 [hep-th]}}.

\bibitem{An:2017hlx}
H.~An, M.~McAneny, A.~K. Ridgway, and M.~B. Wise, ``{Quasi Single Field
  Inflation in the non-perturbative regime},''
  \href{http://dx.doi.org/10.1007/JHEP06(2018)105}{{\em JHEP} {\bfseries 06}
  (2018) 105}, \href{http://arxiv.org/abs/1706.09971}{{\ttfamily
  arXiv:1706.09971 [hep-ph]}}.

\bibitem{Chen:2017ryl}
X.~Chen, Y.~Wang, and Z.-Z. Xianyu, ``{Schwinger-Keldysh Diagrammatics for
  Primordial Perturbations},''
  \href{http://dx.doi.org/10.1088/1475-7516/2017/12/006}{{\em JCAP} {\bfseries
  12} (2017) 006}, \href{http://arxiv.org/abs/1703.10166}{{\ttfamily
  arXiv:1703.10166 [hep-th]}}.

\bibitem{Kumar:2017ecc}
S.~Kumar and R.~Sundrum, ``{Heavy-Lifting of Gauge Theories By Cosmic
  Inflation},'' \href{http://dx.doi.org/10.1007/JHEP05(2018)011}{{\em JHEP}
  {\bfseries 05} (2018) 011}, \href{http://arxiv.org/abs/1711.03988}{{\ttfamily
  arXiv:1711.03988 [hep-ph]}}.

\bibitem{Baumann:2017jvh}
D.~Baumann, G.~Goon, H.~Lee, and G.~L. Pimentel, ``{Partially Massless Fields
  During Inflation},'' \href{http://dx.doi.org/10.1007/JHEP04(2018)140}{{\em
  JHEP} {\bfseries 04} (2018) 140},
  \href{http://arxiv.org/abs/1712.06624}{{\ttfamily arXiv:1712.06624
  [hep-th]}}.

\bibitem{MoradinezhadDizgah:2018ssw}
A.~Moradinezhad~Dizgah, H.~Lee, J.~B. Mu{\~n}oz, and C.~Dvorkin, ``{Galaxy
  Bispectrum from Massive Spinning Particles},''
  \href{http://dx.doi.org/10.1088/1475-7516/2018/05/013}{{\em JCAP} {\bfseries
  05} (2018) 013}, \href{http://arxiv.org/abs/1801.07265}{{\ttfamily
  arXiv:1801.07265 [astro-ph.CO]}}.

\bibitem{Goon:2018fyu}
G.~Goon, K.~Hinterbichler, A.~Joyce, and M.~Trodden, ``{Shapes of gravity:
  Tensor non-Gaussianity and massive spin-2 fields},''
  \href{http://dx.doi.org/10.1007/JHEP10(2019)182}{{\em JHEP} {\bfseries 10}
  (2019) 182}, \href{http://arxiv.org/abs/1812.07571}{{\ttfamily
  arXiv:1812.07571 [hep-th]}}.

\bibitem{Chen:2018xck}
X.~Chen, Y.~Wang, and Z.-Z. Xianyu, ``{Neutrino Signatures in Primordial
  Non-Gaussianities},'' \href{http://dx.doi.org/10.1007/JHEP09(2018)022}{{\em
  JHEP} {\bfseries 09} (2018) 022},
  \href{http://arxiv.org/abs/1805.02656}{{\ttfamily arXiv:1805.02656
  [hep-ph]}}.

\bibitem{Arkani-Hamed:2018kmz}
N.~Arkani-Hamed, D.~Baumann, H.~Lee, and G.~L. Pimentel, ``{The Cosmological
  Bootstrap: Inflationary Correlators from Symmetries and Singularities},''
  \href{http://dx.doi.org/10.1007/JHEP04(2020)105}{{\em JHEP} {\bfseries 04}
  (2020) 105}, \href{http://arxiv.org/abs/1811.00024}{{\ttfamily
  arXiv:1811.00024 [hep-th]}}.

\bibitem{Kumar:2018jxz}
S.~Kumar and R.~Sundrum, ``{Seeing Higher-Dimensional Grand Unification In
  Primordial Non-Gaussianities},''
  \href{http://dx.doi.org/10.1007/JHEP04(2019)120}{{\em JHEP} {\bfseries 04}
  (2019) 120}, \href{http://arxiv.org/abs/1811.11200}{{\ttfamily
  arXiv:1811.11200 [hep-ph]}}.

\bibitem{Wu:2018lmx}
Y.-P. Wu, ``{Higgs as heavy-lifted physics during inflation},''
  \href{http://dx.doi.org/10.1007/JHEP04(2019)125}{{\em JHEP} {\bfseries 04}
  (2019) 125}, \href{http://arxiv.org/abs/1812.10654}{{\ttfamily
  arXiv:1812.10654 [hep-ph]}}.

\bibitem{Dimastrogiovanni:2018uqy}
E.~Dimastrogiovanni, M.~Fasiello, and G.~Tasinato, ``{Probing the inflationary
  particle content: extra spin-2 field},''
  \href{http://dx.doi.org/10.1088/1475-7516/2018/08/016}{{\em JCAP} {\bfseries
  08} (2018) 016}, \href{http://arxiv.org/abs/1806.00850}{{\ttfamily
  arXiv:1806.00850 [astro-ph.CO]}}.

\bibitem{Sleight:2019hfp}
C.~Sleight and M.~Taronna, ``{Bootstrapping Inflationary Correlators in Mellin
  Space},'' \href{http://dx.doi.org/10.1007/JHEP02(2020)098}{{\em JHEP}
  {\bfseries 02} (2020) 098}, \href{http://arxiv.org/abs/1907.01143}{{\ttfamily
  arXiv:1907.01143 [hep-th]}}.

\bibitem{Lu:2019tjj}
S.~Lu, Y.~Wang, and Z.-Z. Xianyu, ``{A Cosmological Higgs Collider},''
  \href{http://dx.doi.org/10.1007/JHEP02(2020)011}{{\em JHEP} {\bfseries 02}
  (2020) 011}, \href{http://arxiv.org/abs/1907.07390}{{\ttfamily
  arXiv:1907.07390 [hep-th]}}.

\bibitem{Hook:2019zxa}
A.~Hook, J.~Huang, and D.~Racco, ``{Searches for other vacua. Part II. A new
  Higgstory at the cosmological collider},''
  \href{http://dx.doi.org/10.1007/JHEP01(2020)105}{{\em JHEP} {\bfseries 01}
  (2020) 105}, \href{http://arxiv.org/abs/1907.10624}{{\ttfamily
  arXiv:1907.10624 [hep-ph]}}.

\bibitem{Hook:2019vcn}
A.~Hook, J.~Huang, and D.~Racco, ``{Minimal signatures of the Standard Model in
  non-Gaussianities},''
  \href{http://dx.doi.org/10.1103/PhysRevD.101.023519}{{\em Phys. Rev. D}
  {\bfseries 101} no.~2, (2020) 023519},
  \href{http://arxiv.org/abs/1908.00019}{{\ttfamily arXiv:1908.00019
  [hep-ph]}}.

\bibitem{Kumar:2019ebj}
S.~Kumar and R.~Sundrum, ``{Cosmological Collider Physics and the Curvaton},''
  \href{http://dx.doi.org/10.1007/JHEP04(2020)077}{{\em JHEP} {\bfseries 04}
  (2020) 077}, \href{http://arxiv.org/abs/1908.11378}{{\ttfamily
  arXiv:1908.11378 [hep-ph]}}.

\bibitem{Wang:2019gbi}
L.-T. Wang and Z.-Z. Xianyu, ``{In Search of Large Signals at the Cosmological
  Collider},'' \href{http://dx.doi.org/10.1007/JHEP02(2020)044}{{\em JHEP}
  {\bfseries 02} (2020) 044}, \href{http://arxiv.org/abs/1910.12876}{{\ttfamily
  arXiv:1910.12876 [hep-ph]}}.

\bibitem{Li:2019ves}
L.~Li, T.~Nakama, C.~M. Sou, Y.~Wang, and S.~Zhou, ``{Gravitational Production
  of Superheavy Dark Matter and Associated Cosmological Signatures},''
  \href{http://dx.doi.org/10.1007/JHEP07(2019)067}{{\em JHEP} {\bfseries 07}
  (2019) 067}, \href{http://arxiv.org/abs/1903.08842}{{\ttfamily
  arXiv:1903.08842 [astro-ph.CO]}}.

\bibitem{Kim:2019wjo}
S.~Kim, T.~Noumi, K.~Takeuchi, and S.~Zhou, ``{Heavy Spinning Particles from
  Signs of Primordial Non-Gaussianities: Beyond the Positivity Bounds},''
  \href{http://dx.doi.org/10.1007/JHEP12(2019)107}{{\em JHEP} {\bfseries 12}
  (2019) 107}, \href{http://arxiv.org/abs/1906.11840}{{\ttfamily
  arXiv:1906.11840 [hep-th]}}.

\bibitem{Baumann:2019oyu}
D.~Baumann, C.~Duaso~Pueyo, A.~Joyce, H.~Lee, and G.~L. Pimentel, ``{The
  cosmological bootstrap: weight-shifting operators and scalar seeds},''
  \href{http://dx.doi.org/10.1007/JHEP12(2020)204}{{\em JHEP} {\bfseries 12}
  (2020) 204}, \href{http://arxiv.org/abs/1910.14051}{{\ttfamily
  arXiv:1910.14051 [hep-th]}}.

\bibitem{Alexander:2019vtb}
S.~Alexander, S.~J. Gates, L.~Jenks, K.~Koutrolikos, and E.~McDonough,
  ``{Higher Spin Supersymmetry at the Cosmological Collider: Sculpting SUSY
  Rilles in the CMB},'' \href{http://dx.doi.org/10.1007/JHEP10(2019)156}{{\em
  JHEP} {\bfseries 10} (2019) 156},
  \href{http://arxiv.org/abs/1907.05829}{{\ttfamily arXiv:1907.05829
  [hep-th]}}.

\bibitem{Bodas:2020yho}
A.~Bodas, S.~Kumar, and R.~Sundrum, ``{The Scalar Chemical Potential in
  Cosmological Collider Physics},''
  \href{http://dx.doi.org/10.1007/JHEP02(2021)079}{{\em JHEP} {\bfseries 02}
  (2021) 079}, \href{http://arxiv.org/abs/2010.04727}{{\ttfamily
  arXiv:2010.04727 [hep-ph]}}.

\bibitem{Kogai:2020vzz}
K.~Kogai, K.~Akitsu, F.~Schmidt, and Y.~Urakawa, ``{Galaxy imaging surveys as
  spin-sensitive detector for cosmological colliders},''
  \href{http://dx.doi.org/10.1088/1475-7516/2021/03/060}{{\em JCAP} {\bfseries
  03} (2021) 060}, \href{http://arxiv.org/abs/2009.05517}{{\ttfamily
  arXiv:2009.05517 [astro-ph.CO]}}.

\bibitem{Aoki:2020zbj}
S.~Aoki and M.~Yamaguchi, ``{Disentangling mass spectra of multiple fields in
  cosmological collider},''
  \href{http://dx.doi.org/10.1007/JHEP04(2021)127}{{\em JHEP} {\bfseries 04}
  (2021) 127}, \href{http://arxiv.org/abs/2012.13667}{{\ttfamily
  arXiv:2012.13667 [hep-th]}}.

\bibitem{Lu:2021gso}
S.~Lu, ``{Axion isocurvature collider},''
  \href{http://dx.doi.org/10.1007/JHEP04(2022)157}{{\em JHEP} {\bfseries 04}
  (2022) 157}, \href{http://arxiv.org/abs/2103.05958}{{\ttfamily
  arXiv:2103.05958 [hep-th]}}.

\bibitem{Wang:2021qez}
L.-T. Wang, Z.-Z. Xianyu, and Y.-M. Zhong, ``{Precision calculation of
  inflation correlators at one loop},''
  \href{http://dx.doi.org/10.1007/JHEP02(2022)085}{{\em JHEP} {\bfseries 02}
  (2022) 085}, \href{http://arxiv.org/abs/2109.14635}{{\ttfamily
  arXiv:2109.14635 [hep-ph]}}.

\bibitem{Lu:2021wxu}
Q.~Lu, M.~Reece, and Z.-Z. Xianyu, ``{Missing scalars at the cosmological
  collider},'' \href{http://dx.doi.org/10.1007/JHEP12(2021)098}{{\em JHEP}
  {\bfseries 12} (2021) 098}, \href{http://arxiv.org/abs/2108.11385}{{\ttfamily
  arXiv:2108.11385 [hep-ph]}}.

\bibitem{Cabass:2021fnw}
G.~Cabass, E.~Pajer, D.~Stefanyszyn, and J.~Supe{\l}, ``{Bootstrapping large
  graviton non-Gaussianities},''
  \href{http://dx.doi.org/10.1007/JHEP05(2022)077}{{\em JHEP} {\bfseries 05}
  (2022) 077}, \href{http://arxiv.org/abs/2109.10189}{{\ttfamily
  arXiv:2109.10189 [hep-th]}}.

\bibitem{Cabass:2021iii}
G.~Cabass, ``{Zoology of graviton non-Gaussianities},''
  \href{http://dx.doi.org/10.1088/1475-7516/2021/12/001}{{\em JCAP} {\bfseries
  12} no.~12, (2021) 001}, \href{http://arxiv.org/abs/2103.09816}{{\ttfamily
  arXiv:2103.09816 [hep-th]}}.

\bibitem{Dimastrogiovanni:2021cif}
E.~Dimastrogiovanni, M.~Fasiello, and A.~E. G{\"u}mr{\"u}kc{\"u}ogl{\"u},
  ``{Spinning guest fields during inflation: leftover signatures},''
  \href{http://dx.doi.org/10.1088/1475-7516/2021/11/047}{{\em JCAP} {\bfseries
  11} no.~11, (2021) 047}, \href{http://arxiv.org/abs/2108.06722}{{\ttfamily
  arXiv:2108.06722 [astro-ph.CO]}}.

\bibitem{Tong:2021wai}
X.~Tong, Y.~Wang, and Y.~Zhu, ``{Cutting rule for cosmological collider
  signals: a bulk evolution perspective},''
  \href{http://dx.doi.org/10.1007/JHEP03(2022)181}{{\em JHEP} {\bfseries 03}
  (2022) 181}, \href{http://arxiv.org/abs/2112.03448}{{\ttfamily
  arXiv:2112.03448 [hep-th]}}.

\bibitem{Cui:2021iie}
Y.~Cui and Z.-Z. Xianyu, ``{Probing Leptogenesis with the Cosmological
  Collider},'' \href{http://dx.doi.org/10.1103/PhysRevLett.129.111301}{{\em
  Phys. Rev. Lett.} {\bfseries 129} no.~11, (2022) 111301},
  \href{http://arxiv.org/abs/2112.10793}{{\ttfamily arXiv:2112.10793
  [hep-ph]}}.

\bibitem{Tong:2022cdz}
X.~Tong and Z.-Z. Xianyu, ``{Large spin-2 signals at the cosmological
  collider},'' \href{http://dx.doi.org/10.1007/JHEP10(2022)194}{{\em JHEP}
  {\bfseries 10} (2022) 194}, \href{http://arxiv.org/abs/2203.06349}{{\ttfamily
  arXiv:2203.06349 [hep-ph]}}.

\bibitem{Pimentel:2022fsc}
G.~L. Pimentel and D.-G. Wang, ``{Boostless cosmological collider bootstrap},''
  \href{http://dx.doi.org/10.1007/JHEP10(2022)177}{{\em JHEP} {\bfseries 10}
  (2022) 177}, \href{http://arxiv.org/abs/2205.00013}{{\ttfamily
  arXiv:2205.00013 [hep-th]}}.

\bibitem{Chen:2022vzh}
X.~Chen, R.~Ebadi, and S.~Kumar, ``{Classical cosmological collider physics and
  primordial features},''
  \href{http://dx.doi.org/10.1088/1475-7516/2022/08/083}{{\em JCAP} {\bfseries
  08} (2022) 083}, \href{http://arxiv.org/abs/2205.01107}{{\ttfamily
  arXiv:2205.01107 [hep-ph]}}.

\bibitem{Jazayeri:2022kjy}
S.~Jazayeri and S.~Renaux-Petel, ``{Cosmological bootstrap in slow motion},''
  \href{http://dx.doi.org/10.1007/JHEP12(2022)137}{{\em JHEP} {\bfseries 12}
  (2022) 137}, \href{http://arxiv.org/abs/2205.10340}{{\ttfamily
  arXiv:2205.10340 [hep-th]}}.

\bibitem{AnilKumar:2022flx}
N.~Anil~Kumar, G.~Sato-Polito, M.~Kamionkowski, and S.~C. Hotinli,
  ``{Primordial trispectrum from kinetic Sunyaev-Zel\textquoteright{}dovich
  tomography},'' \href{http://dx.doi.org/10.1103/PhysRevD.106.063533}{{\em
  Phys. Rev. D} {\bfseries 106} no.~6, (2022) 063533},
  \href{http://arxiv.org/abs/2205.03423}{{\ttfamily arXiv:2205.03423
  [astro-ph.CO]}}.

\bibitem{Maru:2022bhr}
N.~Maru and A.~Okawa, ``{Cosmological collider signals of non-Gaussianity from
  higgs boson in GUT},''
  \href{http://dx.doi.org/10.1142/S0217751X23500756}{{\em Int. J. Mod. Phys. A}
  {\bfseries 38} no.~14, (2023) 2350075},
  \href{http://arxiv.org/abs/2206.06651}{{\ttfamily arXiv:2206.06651
  [hep-ph]}}.

\bibitem{Qin:2022fbv}
Z.~Qin and Z.-Z. Xianyu, ``{Helical inflation correlators: partial
  Mellin-Barnes and bootstrap equations},''
  \href{http://dx.doi.org/10.1007/JHEP04(2023)059}{{\em JHEP} {\bfseries 04}
  (2023) 059}, \href{http://arxiv.org/abs/2208.13790}{{\ttfamily
  arXiv:2208.13790 [hep-th]}}.

\bibitem{Cabass:2022jda}
G.~Cabass, D.~Stefanyszyn, J.~Supe{\l}, and A.~Thavanesan, ``{On graviton
  non-Gaussianities in the Effective Field Theory of Inflation},''
  \href{http://dx.doi.org/10.1007/JHEP10(2022)154}{{\em JHEP} {\bfseries 10}
  (2022) 154}, \href{http://arxiv.org/abs/2209.00677}{{\ttfamily
  arXiv:2209.00677 [hep-th]}}.

\bibitem{Niu:2022quw}
X.~Niu, M.~H. Rahat, K.~Srinivasan, and W.~Xue, ``{Gravitational wave probes of
  massive gauge bosons at the cosmological collider},''
  \href{http://dx.doi.org/10.1088/1475-7516/2023/02/013}{{\em JCAP} {\bfseries
  02} (2023) 013}, \href{http://arxiv.org/abs/2211.14331}{{\ttfamily
  arXiv:2211.14331 [hep-ph]}}.

\bibitem{Chen:2023txq}
X.~Chen, J.~Fan, and L.~Li, ``{New inflationary probes of axion dark matter},''
  \href{http://arxiv.org/abs/2303.03406}{{\ttfamily arXiv:2303.03406
  [hep-ph]}}.

\bibitem{Qin:2023bjk}
Z.~Qin and Z.-Z. Xianyu, ``{Inflation correlators at the one-loop order:
  nonanalyticity, factorization, cutting rule, and OPE},''
  \href{http://dx.doi.org/10.1007/JHEP09(2023)116}{{\em JHEP} {\bfseries 09}
  (2023) 116}, \href{http://arxiv.org/abs/2304.13295}{{\ttfamily
  arXiv:2304.13295 [hep-th]}}.

\bibitem{Chakraborty:2023qbp}
P.~Chakraborty and J.~Stout, ``{Light Scalars at the Cosmological Collider},''
  \href{http://arxiv.org/abs/2310.01494}{{\ttfamily arXiv:2310.01494
  [hep-th]}}.

\bibitem{Chakraborty:2023eoq}
P.~Chakraborty and J.~Stout, ``{Compact scalars at the cosmological
  collider},'' \href{http://dx.doi.org/10.1007/JHEP03(2024)149}{{\em JHEP}
  {\bfseries 03} (2024) 149}, \href{http://arxiv.org/abs/2311.09219}{{\ttfamily
  arXiv:2311.09219 [hep-th]}}.

\bibitem{Aoki:2023dsl}
S.~Aoki, T.~Noumi, F.~Sano, and M.~Yamaguchi, ``{Analytic Formulae for
  Inflationary Correlators with Dynamical Mass},''
  \href{http://arxiv.org/abs/2312.09642}{{\ttfamily arXiv:2312.09642
  [hep-th]}}.

\bibitem{Craig:2024qgy}
N.~Craig, S.~Kumar, and A.~McCune, ``{An effective cosmological collider},''
  \href{http://dx.doi.org/10.1007/JHEP07(2024)108}{{\em JHEP} {\bfseries 07}
  (2024) 108}, \href{http://arxiv.org/abs/2401.10976}{{\ttfamily
  arXiv:2401.10976 [hep-ph]}}.

\bibitem{Quintin:2024boj}
J.~Quintin, X.~Chen, and R.~Ebadi, ``{Fingerprints of a non-inflationary
  universe from massive fields},''
  \href{http://dx.doi.org/10.1088/1475-7516/2024/09/026}{{\em JCAP} {\bfseries
  09} (2024) 026}, \href{http://arxiv.org/abs/2405.11016}{{\ttfamily
  arXiv:2405.11016 [astro-ph.CO]}}.

\bibitem{Goldstein:2024bky}
S.~Goldstein, O.~H.~E. Philcox, J.~C. Hill, and L.~Hui, ``{Intermediate
  mass-range particles from small scales: Nonperturbative techniques for
  cosmological collider physics from large-scale structure surveys},''
  \href{http://dx.doi.org/10.1103/PhysRevD.110.083516}{{\em Phys. Rev. D}
  {\bfseries 110} no.~8, (2024) 083516},
  \href{http://arxiv.org/abs/2407.08731}{{\ttfamily arXiv:2407.08731
  [astro-ph.CO]}}.

\bibitem{Qin:2024gtr}
Z.~Qin, ``{Cosmological correlators at the loop level},''
  \href{http://dx.doi.org/10.1007/JHEP03(2025)051}{{\em JHEP} {\bfseries 03}
  (2025) 051}, \href{http://arxiv.org/abs/2411.13636}{{\ttfamily
  arXiv:2411.13636 [hep-th]}}.

\bibitem{Hubisz:2024xnj}
J.~Hubisz, S.~J. Lee, H.~Li, and B.~Sambasivam, ``{Cosmological quasiparticles
  and the cosmological collider},''
  \href{http://dx.doi.org/10.1103/PhysRevD.111.023543}{{\em Phys. Rev. D}
  {\bfseries 111} no.~2, (2025) 023543},
  \href{http://arxiv.org/abs/2408.08951}{{\ttfamily arXiv:2408.08951
  [astro-ph.CO]}}.

\bibitem{Liu:2024str}
H.~Liu and Z.-Z. Xianyu, ``{Massive Inflationary Amplitudes: Differential
  Equations and Complete Solutions for General Trees},''
  \href{http://arxiv.org/abs/2412.07843}{{\ttfamily arXiv:2412.07843
  [hep-th]}}.

\bibitem{Stefanyszyn:2024msm}
D.~Stefanyszyn, X.~Tong, and Y.~Zhu, ``{There and Back Again: Mapping and
  Factorizing Cosmological Observables},''
  \href{http://dx.doi.org/10.1103/PhysRevLett.133.221501}{{\em Phys. Rev.
  Lett.} {\bfseries 133} no.~22, (2024) 221501},
  \href{http://arxiv.org/abs/2406.00099}{{\ttfamily arXiv:2406.00099
  [hep-th]}}.

\bibitem{Bodas:2024hih}
A.~Bodas, E.~Broadberry, and R.~Sundrum, ``{Grand unification at the
  cosmological collider with chemical potential},''
  \href{http://dx.doi.org/10.1007/JHEP01(2025)115}{{\em JHEP} {\bfseries 01}
  (2025) 115}, \href{http://arxiv.org/abs/2409.07524}{{\ttfamily
  arXiv:2409.07524 [hep-ph]}}.

\bibitem{Cespedes:2025dnq}
S.~Cespedes and S.~Jazayeri, ``{The massive flat space limit of cosmological
  correlators},'' \href{http://dx.doi.org/10.1007/JHEP07(2025)032}{{\em JHEP}
  {\bfseries 07} (2025) 032}, \href{http://arxiv.org/abs/2501.02119}{{\ttfamily
  arXiv:2501.02119 [hep-th]}}.

\bibitem{Chakraborty:2025myb}
P.~Chakraborty, ``{Primordial Non-Gaussianity from Light Compact Scalars},''
  \href{http://arxiv.org/abs/2501.07672}{{\ttfamily arXiv:2501.07672
  [hep-th]}}.

\bibitem{deRham:2025mjh}
C.~de~Rham, S.~Jazayeri, and A.~J. Tolley, ``{Bispectrum islands: Bootstrap
  bounds on cosmological correlators},''
  \href{http://dx.doi.org/10.1103/q6rq-sj9t}{{\em Phys. Rev. D} {\bfseries 112}
  no.~8, (2025) 083531}, \href{http://arxiv.org/abs/2506.19198}{{\ttfamily
  arXiv:2506.19198 [hep-th]}}.

\bibitem{Zhang:2025nzd}
H.~Zhang, ``{Dimensional Regularization of Bubble Diagrams in de Sitter
  Spacetime},'' \href{http://arxiv.org/abs/2507.19318}{{\ttfamily
  arXiv:2507.19318 [hep-th]}}.

\bibitem{Bodas:2025wuk}
A.~Bodas, E.~Broadberry, R.~Sundrum, and Z.~Xu, ``{Charged Loops at the
  Cosmological Collider with Chemical Potential},''
  \href{http://arxiv.org/abs/2507.22978}{{\ttfamily arXiv:2507.22978
  [hep-ph]}}.

\bibitem{Qin:2025xct}
Z.~Qin, S.~Renaux-Petel, X.~Tong, D.~Werth, and Y.~Zhu, ``{The Exact and
  Approximate Tales of Boost-Breaking Cosmological Correlators},''
  \href{http://arxiv.org/abs/2506.01555}{{\ttfamily arXiv:2506.01555
  [hep-th]}}.

\bibitem{Chakraborty:2025mhh}
P.~Chakraborty, T.~Cohen, D.~Green, and Y.~Huang, ``{A Compact Story of
  Positivity in de Sitter},'' \href{http://arxiv.org/abs/2508.08359}{{\ttfamily
  arXiv:2508.08359 [hep-th]}}.

\bibitem{Aoki:2025uff}
S.~Aoki and A.~Strumia, ``{Testing the arrow of time at the cosmo collider},''
  \href{http://arxiv.org/abs/2510.05204}{{\ttfamily arXiv:2510.05204
  [hep-ph]}}.

\bibitem{Goldberger:1999uk}
W.~D. Goldberger and M.~B. Wise, ``{Modulus stabilization with bulk fields},''
  \href{http://dx.doi.org/10.1103/PhysRevLett.83.4922}{{\em Phys. Rev. Lett.}
  {\bfseries 83} (1999) 4922--4925},
  \href{http://arxiv.org/abs/hep-ph/9907447}{{\ttfamily arXiv:hep-ph/9907447}}.

\bibitem{Goldberger:1999un}
W.~D. Goldberger and M.~B. Wise, ``{Phenomenology of a stabilized modulus},''
  \href{http://dx.doi.org/10.1016/S0370-2693(00)00099-X}{{\em Phys. Lett. B}
  {\bfseries 475} (2000) 275--279},
  \href{http://arxiv.org/abs/hep-ph/9911457}{{\ttfamily arXiv:hep-ph/9911457}}.

\bibitem{Lyth:2001nq}
D.~H. Lyth and D.~Wands, ``{Generating the curvature perturbation without an
  inflaton},'' \href{http://dx.doi.org/10.1016/S0370-2693(01)01366-1}{{\em
  Phys. Lett. B} {\bfseries 524} (2002) 5--14},
  \href{http://arxiv.org/abs/hep-ph/0110002}{{\ttfamily arXiv:hep-ph/0110002}}.

\bibitem{Moroi:2001ct}
T.~Moroi and T.~Takahashi, ``{Effects of cosmological moduli fields on cosmic
  microwave background},''
  \href{http://dx.doi.org/10.1016/S0370-2693(01)01295-3}{{\em Phys. Lett. B}
  {\bfseries 522} (2001) 215--221},
  \href{http://arxiv.org/abs/hep-ph/0110096}{{\ttfamily arXiv:hep-ph/0110096}}.
  [Erratum: Phys.Lett.B 539, 303--303 (2002)].

\bibitem{Enqvist:2001zp}
K.~Enqvist and M.~S. Sloth, ``{Adiabatic CMB perturbations in pre - big bang
  string cosmology},''
  \href{http://dx.doi.org/10.1016/S0550-3213(02)00043-3}{{\em Nucl. Phys. B}
  {\bfseries 626} (2002) 395--409},
  \href{http://arxiv.org/abs/hep-ph/0109214}{{\ttfamily arXiv:hep-ph/0109214}}.

\bibitem{Maldacena:1997re}
J.~M. Maldacena, ``{The Large $N$ limit of superconformal field theories and
  supergravity},'' \href{http://dx.doi.org/10.4310/ATMP.1998.v2.n2.a1}{{\em
  Adv. Theor. Math. Phys.} {\bfseries 2} (1998) 231--252},
  \href{http://arxiv.org/abs/hep-th/9711200}{{\ttfamily arXiv:hep-th/9711200}}.

\bibitem{Witten:1998qj}
E.~Witten, ``{Anti de Sitter space and holography},''
  \href{http://dx.doi.org/10.4310/ATMP.1998.v2.n2.a2}{{\em Adv. Theor. Math.
  Phys.} {\bfseries 2} (1998) 253--291},
  \href{http://arxiv.org/abs/hep-th/9802150}{{\ttfamily arXiv:hep-th/9802150}}.

\bibitem{Gubser:1998bc}
S.~S. Gubser, I.~R. Klebanov, and A.~M. Polyakov, ``{Gauge theory correlators
  from noncritical string theory},''
  \href{http://dx.doi.org/10.1016/S0370-2693(98)00377-3}{{\em Phys. Lett. B}
  {\bfseries 428} (1998) 105--114},
  \href{http://arxiv.org/abs/hep-th/9802109}{{\ttfamily arXiv:hep-th/9802109}}.

\bibitem{Gubser:1999vj}
S.~S. Gubser, ``{AdS / CFT and gravity},''
  \href{http://dx.doi.org/10.1103/PhysRevD.63.084017}{{\em Phys. Rev. D}
  {\bfseries 63} (2001) 084017},
  \href{http://arxiv.org/abs/hep-th/9912001}{{\ttfamily arXiv:hep-th/9912001}}.

\bibitem{Arkani-Hamed:2000ijo}
N.~Arkani-Hamed, M.~Porrati, and L.~Randall, ``{Holography and
  phenomenology},'' \href{http://dx.doi.org/10.1088/1126-6708/2001/08/017}{{\em
  JHEP} {\bfseries 08} (2001) 017},
  \href{http://arxiv.org/abs/hep-th/0012148}{{\ttfamily arXiv:hep-th/0012148}}.

\bibitem{Rattazzi:2000hs}
R.~Rattazzi and A.~Zaffaroni, ``{Comments on the holographic picture of the
  Randall-Sundrum model},''
  \href{http://dx.doi.org/10.1088/1126-6708/2001/04/021}{{\em JHEP} {\bfseries
  04} (2001) 021}, \href{http://arxiv.org/abs/hep-th/0012248}{{\ttfamily
  arXiv:hep-th/0012248}}.

\bibitem{tHooft:1973alw}
G.~'t~Hooft, ``{A Planar Diagram Theory for Strong Interactions},''
  \href{http://dx.doi.org/10.1016/0550-3213(74)90154-0}{{\em Nucl. Phys. B}
  {\bfseries 72} (1974) 461}.

\bibitem{Witten:1979kh}
E.~Witten, ``{Baryons in the 1/n Expansion},''
  \href{http://dx.doi.org/10.1016/0550-3213(79)90232-3}{{\em Nucl. Phys. B}
  {\bfseries 160} (1979) 57--115}.

\bibitem{Han:1998sg}
T.~Han, J.~D. Lykken, and R.-J. Zhang, ``{On Kaluza-Klein states from large
  extra dimensions},'' \href{http://dx.doi.org/10.1103/PhysRevD.59.105006}{{\em
  Phys. Rev. D} {\bfseries 59} (1999) 105006},
  \href{http://arxiv.org/abs/hep-ph/9811350}{{\ttfamily arXiv:hep-ph/9811350}}.

\bibitem{Sohn:2024xzd}
W.~Sohn, D.-G. Wang, J.~R. Fergusson, and E.~P.~S. Shellard, ``{Searching for
  cosmological collider in the Planck CMB data},''
  \href{http://dx.doi.org/10.1088/1475-7516/2024/09/016}{{\em JCAP} {\bfseries
  09} (2024) 016}, \href{http://arxiv.org/abs/2404.07203}{{\ttfamily
  arXiv:2404.07203 [astro-ph.CO]}}.

\bibitem{Cabass:2024wob}
G.~Cabass, O.~H.~E. Philcox, M.~M. Ivanov, K.~Akitsu, S.-F. Chen,
  M.~Simonovi{\'c}, and M.~Zaldarriaga, ``{BOSS constraints on massive
  particles during inflation: The cosmological collider in action},''
  \href{http://dx.doi.org/10.1103/PhysRevD.111.063510}{{\em Phys. Rev. D}
  {\bfseries 111} no.~6, (2025) 063510},
  \href{http://arxiv.org/abs/2404.01894}{{\ttfamily arXiv:2404.01894
  [astro-ph.CO]}}.

\bibitem{Philcox:2025bvj}
O.~H.~E. Philcox, ``{Searching for inflationary physics with the CMB
  trispectrum. I. Primordial theory and optimal estimators},''
  \href{http://dx.doi.org/10.1103/wkj2-q62t}{{\em Phys. Rev. D} {\bfseries 111}
  no.~12, (2025) 123532}, \href{http://arxiv.org/abs/2502.04434}{{\ttfamily
  arXiv:2502.04434 [astro-ph.CO]}}.

\bibitem{Philcox:2025lrr}
O.~H.~E. Philcox, ``{Searching for inflationary physics with the CMB
  trispectrum. II. Code and validation},''
  \href{http://dx.doi.org/10.1103/bprw-3zj1}{{\em Phys. Rev. D} {\bfseries 111}
  no.~12, (2025) 123533}, \href{http://arxiv.org/abs/2502.05258}{{\ttfamily
  arXiv:2502.05258 [astro-ph.CO]}}.

\bibitem{Philcox:2025wts}
O.~H.~E. Philcox, ``{Searching for inflationary physics with the CMB
  trispectrum. III. Constraints from Planck},''
  \href{http://dx.doi.org/10.1103/y81z-g7th}{{\em Phys. Rev. D} {\bfseries 111}
  no.~12, (2025) 123534}, \href{http://arxiv.org/abs/2502.06931}{{\ttfamily
  arXiv:2502.06931 [astro-ph.CO]}}.

\bibitem{Fonseca:2012cj}
J.~Fonseca and D.~Wands, ``{Primordial non-Gaussianity from mixed
  inflaton-curvaton perturbations},''
  \href{http://dx.doi.org/10.1088/1475-7516/2012/06/028}{{\em JCAP} {\bfseries
  06} (2012) 028}, \href{http://arxiv.org/abs/1204.3443}{{\ttfamily
  arXiv:1204.3443 [astro-ph.CO]}}.

\bibitem{Dimopoulos:2003az}
K.~Dimopoulos, D.~H. Lyth, A.~Notari, and A.~Riotto, ``{The Curvaton as a
  pseudoNambu-Goldstone boson},''
  \href{http://dx.doi.org/10.1088/1126-6708/2003/07/053}{{\em JHEP} {\bfseries
  07} (2003) 053}, \href{http://arxiv.org/abs/hep-ph/0304050}{{\ttfamily
  arXiv:hep-ph/0304050}}.

\bibitem{Kasuya:2009up}
S.~Kasuya and M.~Kawasaki, ``{Axion isocurvature fluctuations with extremely
  blue spectrum},'' \href{http://dx.doi.org/10.1103/PhysRevD.80.023516}{{\em
  Phys. Rev. D} {\bfseries 80} (2009) 023516},
  \href{http://arxiv.org/abs/0904.3800}{{\ttfamily arXiv:0904.3800
  [astro-ph.CO]}}.

\bibitem{Kawasaki:2012wr}
M.~Kawasaki, N.~Kitajima, and T.~T. Yanagida, ``{Primordial black hole
  formation from an axionlike curvaton model},''
  \href{http://dx.doi.org/10.1103/PhysRevD.87.063519}{{\em Phys. Rev. D}
  {\bfseries 87} no.~6, (2013) 063519},
  \href{http://arxiv.org/abs/1207.2550}{{\ttfamily arXiv:1207.2550 [hep-ph]}}.

\bibitem{Mazumdar:2010sa}
A.~Mazumdar and J.~Rocher, ``{Particle physics models of inflation and curvaton
  scenarios},'' \href{http://dx.doi.org/10.1016/j.physrep.2010.08.001}{{\em
  Phys. Rept.} {\bfseries 497} (2011) 85--215},
  \href{http://arxiv.org/abs/1001.0993}{{\ttfamily arXiv:1001.0993 [hep-ph]}}.

\bibitem{Lodman:2023yrc}
J.~Lodman, Q.~Lu, and L.~Randall, ``{Savior curvatons and large
  non-Gaussianity},'' \href{http://dx.doi.org/10.1007/JHEP11(2023)218}{{\em
  JHEP} {\bfseries 11} (2023) 218},
  \href{http://arxiv.org/abs/2306.13128}{{\ttfamily arXiv:2306.13128
  [astro-ph.CO]}}.

\bibitem{Lyth:2002my}
D.~H. Lyth, C.~Ungarelli, and D.~Wands, ``{The Primordial density perturbation
  in the curvaton scenario},''
  \href{http://dx.doi.org/10.1103/PhysRevD.67.023503}{{\em Phys. Rev. D}
  {\bfseries 67} (2003) 023503},
  \href{http://arxiv.org/abs/astro-ph/0208055}{{\ttfamily
  arXiv:astro-ph/0208055}}.

\bibitem{Planck:2019kim}
{\bfseries Planck} Collaboration, Y.~Akrami {\em et~al.}, ``{Planck 2018
  results. IX. Constraints on primordial non-Gaussianity},''
  \href{http://dx.doi.org/10.1051/0004-6361/201935891}{{\em Astron. Astrophys.}
  {\bfseries 641} (2020) A9}, \href{http://arxiv.org/abs/1905.05697}{{\ttfamily
  arXiv:1905.05697 [astro-ph.CO]}}.

\bibitem{ACT:2025fju}
{\bfseries ACT} Collaboration, T.~Louis {\em et~al.}, ``{The Atacama Cosmology
  Telescope: DR6 Power Spectra, Likelihoods and $\Lambda$CDM Parameters},''
  \href{http://arxiv.org/abs/2503.14452}{{\ttfamily arXiv:2503.14452
  [astro-ph.CO]}}.

\bibitem{SPHEREx:2014bgr}
{\bfseries SPHEREx} Collaboration, O.~Dor{\'e} {\em et~al.}, ``{Cosmology with
  the SPHEREX All-Sky Spectral Survey},''
  \href{http://arxiv.org/abs/1412.4872}{{\ttfamily arXiv:1412.4872
  [astro-ph.CO]}}.

\bibitem{Csaki:1999mp}
C.~Csaki, M.~Graesser, L.~Randall, and J.~Terning, ``{Cosmology of brane models
  with radion stabilization},''
  \href{http://dx.doi.org/10.1103/PhysRevD.62.045015}{{\em Phys. Rev. D}
  {\bfseries 62} (2000) 045015},
  \href{http://arxiv.org/abs/hep-ph/9911406}{{\ttfamily arXiv:hep-ph/9911406}}.

\bibitem{Csaki:1999jh}
C.~Csaki, M.~Graesser, C.~F. Kolda, and J.~Terning, ``{Cosmology of one extra
  dimension with localized gravity},''
  \href{http://dx.doi.org/10.1016/S0370-2693(99)00896-5}{{\em Phys. Lett. B}
  {\bfseries 462} (1999) 34--40},
  \href{http://arxiv.org/abs/hep-ph/9906513}{{\ttfamily arXiv:hep-ph/9906513}}.

\bibitem{Binetruy:1999hy}
P.~Binetruy, C.~Deffayet, U.~Ellwanger, and D.~Langlois, ``{Brane cosmological
  evolution in a bulk with cosmological constant},''
  \href{http://dx.doi.org/10.1016/S0370-2693(00)00204-5}{{\em Phys. Lett. B}
  {\bfseries 477} (2000) 285--291},
  \href{http://arxiv.org/abs/hep-th/9910219}{{\ttfamily arXiv:hep-th/9910219}}.

\bibitem{Lukas:1998qs}
A.~Lukas, B.~A. Ovrut, and D.~Waldram, ``{Cosmological solutions of
  Horava-Witten theory},''
  \href{http://dx.doi.org/10.1103/PhysRevD.60.086001}{{\em Phys. Rev. D}
  {\bfseries 60} (1999) 086001},
  \href{http://arxiv.org/abs/hep-th/9806022}{{\ttfamily arXiv:hep-th/9806022}}.

\bibitem{Lukas:1999yn}
A.~Lukas, B.~A. Ovrut, and D.~Waldram, ``{Boundary inflation},''
  \href{http://dx.doi.org/10.1103/PhysRevD.61.023506}{{\em Phys. Rev. D}
  {\bfseries 61} (2000) 023506},
  \href{http://arxiv.org/abs/hep-th/9902071}{{\ttfamily arXiv:hep-th/9902071}}.

\bibitem{Nihei:1999mt}
T.~Nihei, ``{Inflation in the five-dimensional universe with an orbifold extra
  dimension},'' \href{http://dx.doi.org/10.1016/S0370-2693(99)01085-0}{{\em
  Phys. Lett. B} {\bfseries 465} (1999) 81--85},
  \href{http://arxiv.org/abs/hep-ph/9905487}{{\ttfamily arXiv:hep-ph/9905487}}.

\bibitem{Maartens:1999hf}
R.~Maartens, D.~Wands, B.~A. Bassett, and I.~Heard, ``{Chaotic inflation on the
  brane},'' \href{http://dx.doi.org/10.1103/PhysRevD.62.041301}{{\em Phys. Rev.
  D} {\bfseries 62} (2000) 041301},
  \href{http://arxiv.org/abs/hep-ph/9912464}{{\ttfamily arXiv:hep-ph/9912464}}.

\bibitem{Karch:2020iit}
A.~Karch and L.~Randall, ``{Geometries with mismatched branes},''
  \href{http://dx.doi.org/10.1007/JHEP09(2020)166}{{\em JHEP} {\bfseries 09}
  (2020) 166}, \href{http://arxiv.org/abs/2006.10061}{{\ttfamily
  arXiv:2006.10061 [hep-th]}}.

\bibitem{Green:2013rd}
D.~Green, M.~Lewandowski, L.~Senatore, E.~Silverstein, and M.~Zaldarriaga,
  ``{Anomalous Dimensions and Non-Gaussianity},''
  \href{http://dx.doi.org/10.1007/JHEP10(2013)171}{{\em JHEP} {\bfseries 10}
  (2013) 171}, \href{http://arxiv.org/abs/1301.2630}{{\ttfamily arXiv:1301.2630
  [hep-th]}}.

\bibitem{Aoki:2023tjm}
S.~Aoki, ``{Continuous spectrum on cosmological collider},''
  \href{http://dx.doi.org/10.1088/1475-7516/2023/04/002}{{\em JCAP} {\bfseries
  04} (2023) 002}, \href{http://arxiv.org/abs/2301.07920}{{\ttfamily
  arXiv:2301.07920 [hep-th]}}.

\bibitem{Pimentel:2025rds}
G.~L. Pimentel and C.~Yang, ``{Strongly Coupled Sectors in Inflation: Gapless
  Theories and Unparticles},''
  \href{http://arxiv.org/abs/2503.17840}{{\ttfamily arXiv:2503.17840
  [hep-th]}}.

\bibitem{Dvali:1998pa}
G.~R. Dvali and S.~H.~H. Tye, ``{Brane inflation},''
  \href{http://dx.doi.org/10.1016/S0370-2693(99)00132-X}{{\em Phys. Lett. B}
  {\bfseries 450} (1999) 72--82},
  \href{http://arxiv.org/abs/hep-ph/9812483}{{\ttfamily arXiv:hep-ph/9812483}}.

\bibitem{Kaloper:1999sm}
N.~Kaloper, ``{Bent domain walls as brane worlds},''
  \href{http://dx.doi.org/10.1103/PhysRevD.60.123506}{{\em Phys. Rev. D}
  {\bfseries 60} (1999) 123506},
  \href{http://arxiv.org/abs/hep-th/9905210}{{\ttfamily arXiv:hep-th/9905210}}.

\bibitem{Arkani-Hamed:1999fet}
N.~Arkani-Hamed, S.~Dimopoulos, N.~Kaloper, and J.~March-Russell, ``{Rapid
  asymmetric inflation and early cosmology in theories with submillimeter
  dimensions},'' \href{http://dx.doi.org/10.1016/S0550-3213(99)00667-7}{{\em
  Nucl. Phys. B} {\bfseries 567} (2000) 189--228},
  \href{http://arxiv.org/abs/hep-ph/9903224}{{\ttfamily arXiv:hep-ph/9903224}}.

\bibitem{Burgess:2001fx}
C.~P. Burgess, M.~Majumdar, D.~Nolte, F.~Quevedo, G.~Rajesh, and R.-J. Zhang,
  ``{The Inflationary brane anti-brane universe},''
  \href{http://dx.doi.org/10.1088/1126-6708/2001/07/047}{{\em JHEP} {\bfseries
  07} (2001) 047}, \href{http://arxiv.org/abs/hep-th/0105204}{{\ttfamily
  arXiv:hep-th/0105204}}.

\bibitem{Garcia-Bellido:2001lbk}
J.~Garcia-Bellido, R.~Rabadan, and F.~Zamora, ``{Inflationary scenarios from
  branes at angles},''
  \href{http://dx.doi.org/10.1088/1126-6708/2002/01/036}{{\em JHEP} {\bfseries
  01} (2002) 036}, \href{http://arxiv.org/abs/hep-th/0112147}{{\ttfamily
  arXiv:hep-th/0112147}}.

\bibitem{Jones:2002cv}
N.~T. Jones, H.~Stoica, and S.~H.~H. Tye, ``{Brane interaction as the origin of
  inflation},'' \href{http://dx.doi.org/10.1088/1126-6708/2002/07/051}{{\em
  JHEP} {\bfseries 07} (2002) 051},
  \href{http://arxiv.org/abs/hep-th/0203163}{{\ttfamily arXiv:hep-th/0203163}}.

\bibitem{Kachru:2003sx}
S.~Kachru, R.~Kallosh, A.~D. Linde, J.~M. Maldacena, L.~P. McAllister, and
  S.~P. Trivedi, ``{Towards inflation in string theory},''
  \href{http://dx.doi.org/10.1088/1475-7516/2003/10/013}{{\em JCAP} {\bfseries
  10} (2003) 013}, \href{http://arxiv.org/abs/hep-th/0308055}{{\ttfamily
  arXiv:hep-th/0308055}}.

\bibitem{Silverstein:2003hf}
E.~Silverstein and D.~Tong, ``{Scalar speed limits and cosmology: Acceleration
  from D-cceleration},''
  \href{http://dx.doi.org/10.1103/PhysRevD.70.103505}{{\em Phys. Rev. D}
  {\bfseries 70} (2004) 103505},
  \href{http://arxiv.org/abs/hep-th/0310221}{{\ttfamily arXiv:hep-th/0310221}}.

\bibitem{Alishahiha:2004eh}
M.~Alishahiha, E.~Silverstein, and D.~Tong, ``{DBI in the sky},''
  \href{http://dx.doi.org/10.1103/PhysRevD.70.123505}{{\em Phys. Rev. D}
  {\bfseries 70} (2004) 123505},
  \href{http://arxiv.org/abs/hep-th/0404084}{{\ttfamily arXiv:hep-th/0404084}}.

\bibitem{DeWolfe:1999cp}
O.~DeWolfe, D.~Z. Freedman, S.~S. Gubser, and A.~Karch, ``{Modeling the
  fifth-dimension with scalars and gravity},''
  \href{http://dx.doi.org/10.1103/PhysRevD.62.046008}{{\em Phys. Rev. D}
  {\bfseries 62} (2000) 046008},
  \href{http://arxiv.org/abs/hep-th/9909134}{{\ttfamily arXiv:hep-th/9909134}}.

\bibitem{Chacko:2001em}
Z.~Chacko and P.~J. Fox, ``{Wave function of the radion in the dS and AdS brane
  worlds},'' \href{http://dx.doi.org/10.1103/PhysRevD.64.024015}{{\em Phys.
  Rev. D} {\bfseries 64} (2001) 024015},
  \href{http://arxiv.org/abs/hep-th/0102023}{{\ttfamily arXiv:hep-th/0102023}}.

\bibitem{Higuchi:1986py}
A.~Higuchi, ``{Forbidden Mass Range for Spin-2 Field Theory in De Sitter
  Space-time},'' \href{http://dx.doi.org/10.1016/0550-3213(87)90691-2}{{\em
  Nucl. Phys. B} {\bfseries 282} (1987) 397--436}.

\bibitem{Chacko:2013dra}
Z.~Chacko, R.~K. Mishra, and D.~Stolarski, ``{Dynamics of a Stabilized Radion
  and Duality},'' \href{http://dx.doi.org/10.1007/JHEP09(2013)121}{{\em JHEP}
  {\bfseries 09} (2013) 121}, \href{http://arxiv.org/abs/1304.1795}{{\ttfamily
  arXiv:1304.1795 [hep-ph]}}.

\bibitem{Giudice:2000av}
G.~F. Giudice, R.~Rattazzi, and J.~D. Wells, ``{Graviscalars from higher
  dimensional metrics and curvature Higgs mixing},''
  \href{http://dx.doi.org/10.1016/S0550-3213(00)00686-6}{{\em Nucl. Phys. B}
  {\bfseries 595} (2001) 250--276},
  \href{http://arxiv.org/abs/hep-ph/0002178}{{\ttfamily arXiv:hep-ph/0002178}}.

\bibitem{Csaki:2000zn}
C.~Csaki, M.~L. Graesser, and G.~D. Kribs, ``{Radion dynamics and electroweak
  physics},'' \href{http://dx.doi.org/10.1103/PhysRevD.63.065002}{{\em Phys.
  Rev. D} {\bfseries 63} (2001) 065002},
  \href{http://arxiv.org/abs/hep-th/0008151}{{\ttfamily arXiv:hep-th/0008151}}.

\bibitem{Lust:2025vyz}
S.~L{\"u}st, M.~Nee, and L.~Randall, ``{More Effective RS Field Theory},''
  \href{http://arxiv.org/abs/2510.11771}{{\ttfamily arXiv:2510.11771
  [hep-ph]}}.

\bibitem{Bordin:2019tyb}
L.~Bordin and G.~Cabass, ``{Probing higher-spin fields from inflation with
  higher-order statistics of the CMB},''
  \href{http://dx.doi.org/10.1088/1475-7516/2019/06/050}{{\em JCAP} {\bfseries
  06} (2019) 050}, \href{http://arxiv.org/abs/1902.09519}{{\ttfamily
  arXiv:1902.09519 [astro-ph.CO]}}.

\bibitem{Floss:2022grj}
T.~Fl{\"o}ss, T.~de~Wild, P.~D. Meerburg, and L.~V.~E. Koopmans, ``{The Dark
  Ages' 21-cm trispectrum},''
  \href{http://dx.doi.org/10.1088/1475-7516/2022/06/020}{{\em JCAP} {\bfseries
  06} no.~06, (2022) 020}, \href{http://arxiv.org/abs/2201.08843}{{\ttfamily
  arXiv:2201.08843 [astro-ph.CO]}}.

\bibitem{Coradeschi:2013gda}
F.~Coradeschi, P.~Lodone, D.~Pappadopulo, R.~Rattazzi, and L.~Vitale, ``{A
  naturally light dilaton},''
  \href{http://dx.doi.org/10.1007/JHEP11(2013)057}{{\em JHEP} {\bfseries 11}
  (2013) 057}, \href{http://arxiv.org/abs/1306.4601}{{\ttfamily arXiv:1306.4601
  [hep-th]}}.

\bibitem{Montero:2021otb}
M.~Montero, C.~Vafa, T.~Van~Riet, and V.~Venken, ``{The FL bound and its
  phenomenological implications},''
  \href{http://dx.doi.org/10.1007/JHEP10(2021)009}{{\em JHEP} {\bfseries 10}
  (2021) 009}, \href{http://arxiv.org/abs/2106.07650}{{\ttfamily
  arXiv:2106.07650 [hep-th]}}.

\bibitem{Mishra:2022fic}
R.~K. Mishra, ``{Confinement in de Sitter space and the swampland},''
  \href{http://dx.doi.org/10.1007/JHEP01(2023)002}{{\em JHEP} {\bfseries 01}
  (2023) 002}, \href{http://arxiv.org/abs/2207.12364}{{\ttfamily
  arXiv:2207.12364 [hep-th]}}.

\bibitem{Smith:2015uia}
K.~M. Smith, L.~Senatore, and M.~Zaldarriaga, ``{Optimal analysis of the CMB
  trispectrum},'' \href{http://arxiv.org/abs/1502.00635}{{\ttfamily
  arXiv:1502.00635 [astro-ph.CO]}}.

\bibitem{Wang:2020ioa}
L.-T. Wang and Z.-Z. Xianyu, ``{Gauge Boson Signals at the Cosmological
  Collider},'' \href{http://dx.doi.org/10.1007/JHEP11(2020)082}{{\em JHEP}
  {\bfseries 11} (2020) 082}, \href{http://arxiv.org/abs/2004.02887}{{\ttfamily
  arXiv:2004.02887 [hep-ph]}}.

\bibitem{Pajer:2024ckd}
E.~Pajer, D.-G. Wang, and B.~Zhang, ``{The UV Sensitivity of Axion Monodromy
  Inflation},'' \href{http://arxiv.org/abs/2412.05762}{{\ttfamily
  arXiv:2412.05762 [hep-th]}}.

\end{thebibliography}\endgroup

\end{document}